\documentclass[fleqn,usenatbib]{mnras}

\usepackage{newtxtext,newtxmath}
\usepackage[T1]{fontenc}

\DeclareRobustCommand{\VAN}[3]{#2}
\let\VANthebibliography\thebibliography
\def\thebibliography{\DeclareRobustCommand{\VAN}[3]{##3}\VANthebibliography}

\usepackage{graphicx}	% Including figure files
\usepackage{amsmath}	% Advanced maths commands
\usepackage{subcaption} % For subfigures
\usepackage{upgreek} % For upright Greek letters
\usepackage{xspace} % For correct spaces after new commands
\usepackage{color, colortbl} % For color font during drafting

\newcommand{\cii}{[C\,{\sc ii}]\xspace}
\newcommand{\oiii}{[O\,{\sc iii}]\xspace}
\newcommand{\hii}{H\,{\sc ii}\xspace}
\newcommand{\lya}{Ly{$\rm \alpha$}\xspace}
\newcommand{\ha}{H{$\rm \alpha$}\xspace}

\title[Spatial offsets between stellar and ISM emission in ALPINE]{The ALPINE-ALMA \cii survey: Characterisation of Spatial Offsets in Main-Sequence Galaxies at $z \sim$ 4--6}
\author[M. Killi et al.]{
Meghana Killi,$^{1,2,3,4}$\thanks{E-mail: meghana.killi@nbi.ku.dk}
Michele Ginolfi,$^{5,6}$
Gergö Popping,$^{3}$ % orcid: 0000-0003-1151-4659
Darach Watson,$^{1,2}$
Giovanni Zamorani,$^{7}$
\newauthor Brian C. Lemaux,$^{8,9}$
Seiji Fujimoto,$^{10}$
Andreas Faisst,$^{11}$
Matthieu Bethermin,$^{12,13}$
Michael Romano,$^{14,15}$
\newauthor Yoshinobu Fudamoto,$^{16,17}$
Sandro Bardelli,$^{7}$
Médéric Boquien,$^{18}$
Stefano Carniani,$^{19}$
\newauthor Miroslava Dessauges-Zavadsky,$^{20}$
Carlotta Gruppioni,$^{21}$
Nimish Hathi,$^{22}$
Eduardo Ibar,$^{23}$
Gareth C. Jones,$^{24}$
\newauthor Anton M. Koekemoer,$^{22}$
Ivanna Langan,$^{3,25}$
Hugo Méndez-Hernández,$^{26,27}$
Yuma Sugahara$^{16,17}$
Livia Vallini,$^{7}$
\newauthor Daniela Vergani$^{21}$
\\
$^{1}$Cosmic Dawn Center (DAWN), Jagtvej 128, 2200 Copenhagen N, Denmark\\
$^{2}$Niels Bohr Institute, University of Copenhagen, Lyngbyvej 2, 2100 Copenhagen Ø, Denmark\\
$^{3}$European Southern Observatory, Karl-Schwarzschild-Straße 2, 85748 Garching bei München, Germany\\
$^{4}$Instituto de Estudios Astrofísicos, Facultad de Ingeniería y Ciencias, Universidad Diego Portales, Av. Ejército 441, Santiago 8370191, Chile\\
$^{5}$Dipartimento di Fisica e Astronomia, Università di Firenze, Via G. Sansone 1, 50019, Sesto Fiorentino (Firenze), Italy\\
$^{6}$ INAF - Osservatorio Astrofisico di Arcetri, Largo E. Fermi 5, I-50125, Firenze, Italy\\
$^{7}$INAF – Osservatorio di Astrofisica e Scienza dello Spazio, Via P. Gobetti 93/3, I-40129 Bologna, Italy\\
$^{8}$Department of Physics and Astronomy, University of California, Davis, One Shields Ave., Davis, CA 95616, USA\\ 
$^{9}$Gemini Observatory, NSF’s NOIRLab, 670 N. A’ohoku Place, Hilo, Hawai’i 96720, USA\\
$^{10}$Department of Astronomy, The University of Texas at Austin, 2515 Speedway Blvd Stop C1400, Austin, TX 78712, USA\\
$^{11}$IPAC, California Institute of Technology, 1200 East California Boulevard, Pasadena, CA 91125, USA\\
$^{12}$Aix Marseille Univ, CNRS, CNES, LAM, Marseille, France\\
$^{13}$Université de Strasbourg, CNRS, Observatoire astronomique de Strasbourg, UMR 7550, 67000, Strasbourg, France\\
$^{14}$National Centre for Nuclear Research, ul. Pasteura 7, 02-093 Warsaw, Poland\\
$^{15}$INAF - Osservatorio Astronomico di Padova, Vicolo dell’Osservatorio 5, I-35122, Padova, Italy\\
$^{16}$National Astronomical Observatory of Japan, 2-21-1 Osawa, Mitaka, Tokyo 181-8588, Japan\\
$^{17}$Waseda Research Institute for Science and Engineering, Faculty of Science and Engineering, Waseda University, 3-4-1, Okubo, Shinjuku, Tokyo 169-8555, Japan\\
$^{18}$Centro de Astronomía (CITEVA), Universidad de Antofagasta, Avenida Angamos 601, Antofagasta, Chile\\
$^{19}$Scuola Normale Superiore, Piazza dei Cavalieri 7, I-56126 Pisa, Italy\\
$^{20}$Department of Astronomy, University of Geneva, Chemin Pegasi 51, 1290 Versoix, Switzerland\\
$^{21}$Istituto Nazionale di Astrofisica: Osservatorio di Astrofisica e Scienza dello Spazio di Bologna, Via Gobetti 93/3, 40129 Bologna, Italy\\
$^{22}$Space Telescope Science Institute, Baltimore, MD 21218, USA\\
$^{23}$Instituto de Física y Astronomía, Universidad de Valparaíso, Avda. Gran Bretaña 1111, Valparaíso, Chile\\
$^{24}$Department of Physics, University of Oxford, Denys Wilkinson Building, Keble Road, Oxford OX1 3RH, UK\\
$^{25}$Univ Lyon, Univ Lyon1, Ens de Lyon, CNRS, Centre de Recherche Astrophysique de Lyon (CRAL) UMR5574, F-69230 Saint-Genis-Laval, France\\
$^{26}$Instituto Multidisciplinario de Investigación y Postgrado, Universidad de La Serena, Raúl Bitrán 1305, La Serena, Chile\\
$^{27}$Departamento de Astronomía, Universidad de La Serena, Av. Juan Cisternas 1200 N, La Serena, Chile
}

\date{Accepted XXX. Received YYY; in original form ZZZ}

\pubyear{2023}

\begin{document}
\label{firstpage}
\pagerange{\pageref{firstpage}--\pageref{lastpage}}
\maketitle

\begin{abstract}
Galaxy morphology is shaped by stellar activity, feedback, gas and dust properties, and interactions with surroundings, and can therefore provide insight into these processes. In this paper, we study the spatial offsets between stellar and interstellar medium emission in a sample of 54 main-sequence star-forming galaxies at \(z\sim4\text{--}6\) observed with the Atacama Large Millimeter/submillimeter Array (ALMA) and drawn from the ALMA Large Program to INvestigate C$^+$ at Early times (ALPINE). We find no significant spatial offset for the majority (\(\sim\) 70 percent) of galaxies in the sample among any combination of \cii, far-infrared continuum, optical, and ultraviolet emission. However, a fraction of the sample (\(\sim\) 30 percent) shows offsets larger than the median by more than 3\(\sigma\) significance (compared to the uncertainty on the offsets), especially between \cii and ultraviolet emission. We find that these significant offsets are of the order of \(\sim\)0.5--0.7\,arcsec, corresponding to \(\sim\)3.5--4.5 kiloparsecs. The offsets could be caused by a complex dust geometry, strong feedback from stars and active galactic nuclei, large-scale gas inflow and outflow, or a combination of these phenomena. However, our current analysis does not definitively constrain the origin. Future, higher resolution ALMA and JWST observations may help resolve the ambiguity. Regardless, since there exist at least some galaxies that display such large offsets, galaxy models and spectral energy distribution fitting codes cannot assume co-spatial emission in all main-sequence galaxies, and must take into account that the observed emission across wavelengths may be spatially segregated.
\end{abstract}


%%%%%%%%%%%%%%%%%%%%%%%%%%%%%%%%%%%%%%%%%%%%%%%%%%

%%%%%%%%%%%%%%%%% BODY OF PAPER %%%%%%%%%%%%%%%%%%

\section{Introduction}
\label{sec:intro}

% \tcg{The following is a compilation of collaborator comments - MK}

%\href{https://drive.google.com/drive/folders/1G2yZWlWEuKUbEQddnZVtOUor_qbrwl64?usp=sharing}{link to comments folder}

% The modeling of the distribution is hard to follow. What about a full null test for which many positions are drawn (for each tracer) several times for each source? This would provide the offset expected in the null hypothesis including all the effects. It could also take into account the impact beam elongation on the distribution (it may become no Gaussian, isn't it?). Then, a simple KS test could quantify how unlikely is this null hypothesis. - MR

% quantify correlations via Spearman’s rank test, include Spearman's Rank Correlation Coefficient in each panel as a label; Monte Carlo'ing Spearman tests for each case, i.e., resampling the data using its uncertainties and creating a set of 1000 Spearman correlation coefficients and significances for each relation

% Centroid fitting - fit for HST? - BL - does not work though; try a non-parametric centroid? Or is the goodness of fit generally acceptable for the parametric fit?

% questions about what the “optical” data is - what is this data? observed frame optical or rest-frame optical? If rest-frame, you should mention which data you used. IRAC and where/how did you obtain the data? - YF

% 2 sigma masking - biasing towards peak; don’t do it - try looking at images with 2sig masking applied - do they look very different from no masking images?

% \tcg{The main body of the paper begins here - MK}

%12.9 to 11.4 Gyr ago
The redshift 6 to 4 era corresponds to the period between the end of reionisation of the Universe where the earliest galaxies lived \citep[e.g.][]{Fan2006ObservationalReionization,Dayal2018EarlyEffects, Robertson2022GalaxyTelescope}, and the beginning of cosmic noon where the bulk of the Universe's stellar mass was created \citep[e.g.][]{ForsterSchreiber2020Star-FormingNoon}. This transition period is therefore of utmost interest to trace the evolution of galaxies from first light to the present day.

The most representative galaxies at \(z\sim4\text{--}6\) are those that populate the star-forming main sequence relation (between stellar mass, M$_\star$, and star-formation rate, SFR) at these redshifts \citep[e.g.][]{Noeske2007Galaxies,Speagle2014AZ0-6,Popesso2022TheTimes}, and should hence be ideal to study the physics that led to the eventual creation of modern galaxies. Since observations based only on ultraviolet (UV) emission are limited by dust attenuation \citep{Fudamoto2020TheZ4.4-5.8, Fudamoto2021NormalReionization}, comprehensive studies of main-sequence galaxies and their dust properties require far-infrared (FIR) continuum observations with e.g. the Atacama Large Millimeter/sub-millimeter Array (ALMA). ALMA also allows the study of the cold gas component through observations of bright rest-frame FIR emission lines such as \cii 158$\upmu$m, a major coolant of the interstellar medium \citep[ISM;][] {Hollenbach1999PhotodissociationGalaxies, Wolfire2022PhotodissociationRegions}, generally emitted from multiple gas phases \citep[ionised, neutral, and molecular gas;][]{Carilli2013CoolGalaxies, Vallini2013Far-infraredGalaxies, Vallini2017MolecularEmission, Lagache2018TheGalaxies, Zanella2018TheRedshifts}.

Understanding the physics occurring within galaxies requires the study of both stellar and ISM phases at high-resolution, but it is observationally expensive to conduct high-resolution studies at these redshifts. An alternative is to study the spatial offset between centroids of emission at different wavelengths, which can be done even with low-resolution observations. The presence or absence of spatial offset in a given galaxy can reveal how the stellar and ISM phases evolve and interact. Characterising offsets in a statistical sample of main-sequence galaxies can tell us what is normal among high-$z$ galaxies, and separate the exceptional from the ordinary. We may then study the physics that produces these exceptions, and trace its influence on galaxy evolution.

A few recent studies have found spatial offsets of the order of a few kiloparsecs (kpc) among stellar emission (from \hii regions/ionised diffuse gas traced by rest-frame UV/optical continuum or \oiii), ISM gas emission (from metal-enriched/molecular gas traced by \cii or CO), and ISM dust emission (traced by the FIR continuum) in galaxies at $z \sim 4\text{--}6$ \citep{Hodge2012Evidence4, Willott2015STARGALAXIES, Pentericci2016TracingGalaxies, Carniani2018Kiloparsec-scaleZ=57, Matthee2019ResolvedReionization, Fujimoto2020Structure} and beyond, up to $z \sim 8.5$ \citep{Maiolino2015TheALMA, Laporte2017DustGalaxy, Carniani2017ExtendedALMA, Fujimoto2022JWSTEscape, Schouws2022Significant78, Bowler2022TheGalaxies, Inami2022The6.5}. Several state-of-art zoom-in cosmological simulations have also consistently found spatial displacement of a similar scale between \cii/IR bright regions and \oiii/UV bright regions in the ISM of z\(\sim\)5--6 galaxies \citep[e.g.][]{Arata2019RadiativePhases, Katz2017InterpretingReionization, Behrens2018DustySimulations, Katz2019ProbingJWST, Pallottini2019DeepViews, Sommovigo2020WarmImplications, Pallottini2022ASimulations}. While some predict that offsets should be commonplace as the processes driving them are ubiquitous, others suggest that the offset phenomenon is transient, and therefore observation of offsets should be rare.

In this context, a statistical observational study to identify what fraction of high-$z$, main-sequence galaxies display stellar-ISM spatial offsets is yet to be conducted. This knowledge is important because spectral energy distribution (SED) fitting models often assume an energy balance between emission in the UV and FIR (or at least that the emission in UV/optical is coupled to that in FIR) to derive galaxy properties such as stellar mass (M$_\star$), star-formation rate (SFR), and dust content \citep[e.g.][]{DaCunha2008AGalaxies,Boquien2019CIGALE:Emission}. This assumption may not hold when there is a spatial offset causing a segregation of UV and FIR emission.

In order to conduct a systematic study of spatial offsets, we use the ALPINE-ALMA \cii survey \citep[ALMA Large Programme to INvestigate C+ at Early times;][]{Bethermin2020TheProperties, LeFevre2020The6, Faisst2020TheMeasurements}, which is a statistically significant (see Sec.~\ref{sec:alpine}) sample of main-sequence galaxies at $z \sim 4\text{--}6$. In addition to the FIR properties from ALMA, ALPINE is also covered by a wealth of ancillary data from rest-frame UV to mid-IR, making it an ideal sample to perform this analysis.

We calculate spatial offsets between pairs of emissions, and provide a statistical characterisation of the number, significance, and size of the offsets. We then identify galaxies with significant spatial offsets between stellar and ISM emission, and investigate any potential correlations between offsets and galaxy physical properties, e.g. M$_\star$, SFR, and dust attenuation.

We adopt a Flat $\Lambda$CDM cosmology with $\rm H_0=70\,km\,s^{-1}\,Mpc^{-1}$, $\rm \Omega_M=0.3$, and $\rm \Omega_\Lambda=0.7$. For this cosmology, 1” = 6.27 kpc at \(z=5\), i.e., the mean redshift of our study.

\section{Sample and Data Reduction}

\subsection{Basic properties of the full ALPINE sample}
\label{sec:alpine}
The full ALPINE (Project ID: 2017.1.00428.L; PI: O. Le Fèvre) sample consists of 118 main-sequence star-forming galaxies at \(4.4<z<5.9\), excluding the low-transmission (for \cii) atmospheric window between \(4.6<z<5.1\). The targets were selected using spectroscopic redshifts based on Lyman-$\alpha$ and UV ISM lines \citep{Faisst2020TheMeasurements}, and were drawn from the Cosmic Evolution Survey \citep[COSMOS;][]{Scoville2006TheOverview, Scoville2006COSMOSObservations}, the Extended Chandra Deep Field South \citep[ECDFS;][]{Cardamone2010THEECDF-S}, and the Great Observatories Origins Deep Survey \citep[GOODS;][]{Giavalisco2003TheImaging} fields.

The \cii and FIR continuum data consist of \(\sim\)70 hours of ALMA Band 7 observations conducted in cycles 5 and 6. These emissions trace the metal-enriched cold gas and the dust respectively \citep[e.g.][]{Gruppioni2020The6, Ginolfi2020TheUniverse, Pozzi2021TheUniverse}. The ALMA data-cubes were reduced and calibrated using the standard Common Astronomy Software Applications \citep[CASA;][]{McMullin2007CASAApplications} pipeline. Each cube was continuum-subtracted in the \textit{uv}-plane, and a line search algorithm was then applied to detect the \cii line with SNR > 3.5. ALPINE cubes and images have a pixel scale of 0.15\,arcsec\,pixel$^{-1}$ %, spatial resolution of $\lesssim$1.0\,arcsec, == beam size?
and a 1$\sigma$ sensitivity on the integrated \cii luminosity $\rm L_{\rm [CII]}$ of $0.4 \times 10^8 L_\odot$ assuming a line width of 235\,km\,s$^{-1}$. The smallest circularised beams of galaxies in the dataset are of the order of 0.8\,arcsec, while the largest are of the order of 1.5\,arcsec.%For \cii, the smallest beam is 1."05$\times$0."74, and largest beam is 1."65$\times$0."93. For FIR continuum, smallest beam is 0."95$\times$0."72, and largest beam is 1."41$\times$0."93.

For further details on the data reduction, see \citet{Bethermin2020TheProperties}. The ALMA data products (moment maps and continuum images) are publicly available through the ALPINE Data Release 1 repository\footnote{\href{https://cesam.lam.fr/a2c2s/data_release.php}{https://cesam.lam.fr/a2c2s/data\_release.php}}.

In addition to \cii and FIR continuum images, we use rest-frame UV images from the Hubble Space Telescope (\emph{HST}) taken with the Advanced Camera for Surveys (ACS) F814W filter \citep{Scoville2006TheOverview, Koekemoer2007Processing, Koekemoer2011CANDELS:MOSAICS} to trace the young, massive stellar population. These observations have a 3$\sigma$ depth of \(\sim\)29 mag [AB], with a pixel scale of 0.06\,arcsec\,pixel$^{-1}$, and all \emph{HST} images are registered to Gaia DR2 \citep{Faisst2020TheMeasurements}.

We also include K-band emission from the UltraVISTA survey \citep{McCracken2012UltraVISTA:COSMOS} Data Release 4 \citep{Moneti2023VizieR2019} to trace a slightly older stellar population (compared to that seen with \emph{HST}/F814W). The K-band ($\sim$2.2\,$\upmu$m) emission may come from either the rest-frame optical or the near-UV part of a galaxy's spectrum, depending upon its redshift (between 330\,nm and 400\,nm for ALPINE), but we refer to it as the ``optical'' emission throughout this paper. The spatial resolution for these images is in the range 0.74--0.78\,arcsec, with a seeing of $\sim$ 0.64\,arcsec, and a limiting magnitude of 24.9 [AB] \citep[computed as the 5$\sigma$ limit in a 2.0\,arcsec aperture,][]{Moneti2019TheRelease}. %1$\sigma$ rms in both directions for stars selected with 17.0<$\rm K_s$<19.5 is $\sim$0."08.
The images have a pixel scale of 0.15\,arcsec pixel$^{-1}$ \citep{McCracken2012UltraVISTA:COSMOS}, the same as the ALPINE \cii and FIR continuum images.

Galaxy physical properties were presented by \citet{Faisst2020TheMeasurements}. UV continuum and absorption line properties were obtained from imaging and spectroscopy with \emph{HST}, Keck, and various other instruments, optical lines were inferred from Spitzer photometry, and FIR lines from ALMA. M$_\star$, SFR, light-weighted stellar population age, absolute magnitude, optical dust reddening, and UV continuum slope were obtained via SED fitting with the \texttt{LePhare} code \citep{Arnouts1999MeasuringNorth, Ilbert2006AccurateSurvey, Arnouts2011LePHARE:Estimate}, using synthetic templates based on the \citet{Bruzual2003Stellar2003} stellar population library, tuned to represent galaxies at $4<z<6$. ALPINE galaxies were found to span a range of stellar masses ($\rm log(M_\star/M_\odot)\sim9\text{--}11$) and SFRs ($\rm log(SFR/M_\odot yr^{-1})\sim1\text{--}3$). \ha emission properties including line luminosity and equivalent width were obtained from the Spitzer [3.6 $\upmu$m]--[4.5 $\upmu$m] colour. \ha luminosity was in turn used to derive an estimate of the SFR using the \citet{Kennicutt1998STARSEQUENCE} relation. \cii-ISM velocity offsets were determined by \citet{Cassata2020The6}. For further information on the reduction and properties of ALPINE ancillary data, see \citet{Faisst2020TheMeasurements}.

% For a more in-depth description of the overall ALPINE survey,  the  observations,  data  processing,  and  multi-wavelength analysis see Le Fèvre et al. (2020), Béthermin et al. (2020) and Faisst et al. (2020), respectively.

\subsection{Our sample}
\label{sec:sample}
\begin{table}
\centering
    \begin{tabular}{c c c}
    \hline\hline
    Emission tracer & Number & Fields\\
    \hline
    UV & 54 & COSMOS, GOODS-S, ECDFS\\
    Optical & 45 & COSMOS\\
    \cii & 52 & COSMOS, GOODS-S, ECDFS\\
    FIR continuum & 16 & COSMOS, GOODS-S, ECDFS\\
    \hline
    \end{tabular}
    \caption{Number of galaxies in our sample with each emission tracer observation, and the fields in which the galaxies are located}
    \label{tab:subsamples_obs}
\end{table}

Of the 118 ALPINE galaxies, 75 were detected in \cii emission, and 23 in FIR continuum (21 galaxies have both \cii and FIR continuum detection). \citet{Romano2021The5} performed a morpho-kinematic classification based on \cii emission to identify mergers and multi-component systems and found 23 such merging sources. It is important  to note that while their analysis excludes \emph{major} mergers that can be discerned at our current resolution, there may still be minor or close mergers, satellites, accretion, and clumps at smaller scales. Moreover, as these mergers were identified based mostly on \cii emission, there may still be multiple components in the continuum, optical, or UV emission (see Sec.~\ref{sec:morpho}). For the purpose of this paper, we will exclude the mergers identified as such in \citet{Romano2021The5} (cross-checked with the ``MER'' class in \citet{Jones2021The4.4-5.9}, which is a subset of mergers from \citet{Romano2021The5}) and only consider the remaining population.

This ``non-merging'' ALPINE sample consists of 54 galaxies, of which 52 have \cii detection, and 16 have FIR continuum detection (14 galaxies have both \cii and FIR continuum detection; CANDELS\_GOODSS\_19 and DEIMOS\_COSMOS\_460378 have FIR continuum but no \cii detection). All 54 galaxies have UV observations with \emph{HST}. 45 of these sources, covered by the COSMOS field, have deep UltraVISTA observations in the K-band (the remaining 9 sources in ECDFS and GOODS-S are excluded as they are not detected or barely detected in K-band). 12 galaxies in COSMOS have detections in all four emissions. In Table~\ref{tab:subsamples_obs}, we show the distribution of the final sample, indicating the number of galaxies that have observations in the UV, optical, \cii, and FIR continuum.
% The two galaxies with cont but without cii are CANDELS_GOODSS_19 and DEIMOS_COSMOS_460378

\section{Methods}
In this section, we describe the methods used in this work to calculate centroids of emission in the four emission tracers, and derive spatial offsets among them.

\subsection{UV centroids}
\label{sec:uv_cent}
We use \emph{HST} images taken with the ACS/F814W filter (see Sec.~\ref{sec:alpine}), tracing rest-frame UV emission at the redshift of our targets. The UV centroids are assumed to be the \emph{HST} RA and Dec coordinates from the \citet{Faisst2020TheMeasurements} catalogue. For each galaxy, an astrometric correction is provided in this catalogue as $\delta$RA and $\delta$Dec values to be added to the \emph{HST} coordinates so that the image is aligned with the Gaia DR2 \citep{Mignard2018GaiaGaia-CRF2} catalogue. \citet{Faisst2020TheMeasurements} find an additional scatter of \(\sim\)30\,mas in both RA and Dec after the astrometric correction is applied. We calculate the total UV centroid uncertainty as the sum in quadrature of uncertainty on RA and Dec, which amounts to \(\sim\)40\,mas for all UV images.

\subsection{Centroid fitting and uncertainty}
\label{sec:fitting}
For \cii, FIR continuum, and optical images, we find the centroid of emission and estimate uncertainties in the following way.

\subsubsection{\texorpdfstring{\cii}{[CII]} and FIR continuum}
\label{sec:cii_fitting}
For each galaxy detected in \cii or FIR continuum (or both), we crop the corresponding ALMA moment-0 and/or continuum maps into cutouts of \(6.0\times6.0\)\,arcsec (\(40\times40\) pixels) around the UV centroid position (see Sec.~\ref{sec:uv_cent}). 
To estimate the typical noise level in the image, we calculate the sigma clipped standard deviation of pixel values within an annulus of inner and outer radii of 4.5 and 9.0\,arcsec (30 and 60 pixels). We use this to apply a \(2\sigma\) masking to the cutout image and exclude pixels that are below this significance level. For the pixels with significance \(> 2\sigma\), we fit a two-dimensional, elliptical Gaussian of the form
\begin{equation}
\label{eq:elliptical_gaussian}
    f(x,y) = A e^{- (a(x-x_0)^2 + 2b(x-x_0)(y-y_0) + c(y-y_0)^2)},
\end{equation}
where
\begin{equation}
    a = \frac{\cos^2(\theta)}{2\sigma_x^2} + \frac{\sin^2(\theta)}{2\sigma_y^2},
\end{equation}
\begin{equation}
    b = -\frac{\sin(2\theta)}{4\sigma_x^2} + \frac{\sin(2\theta)}{4\sigma_y^2},
\end{equation}
and
\begin{equation}
    c = \frac{\sin^2(\theta)}{2\sigma_x^2} + \frac{\cos^2(\theta)}{2\sigma_y^2}.
\end{equation}
Here, $x_0$, $y_0$ are the coordinates of the centre, $\sigma_x$, $\sigma_y$ are the Gaussian widths along each dimension, and $\theta$ is the counter-clockwise angle.

We use \textsc{scipy.optimize.curve\_fit} \citep{Virtanen2020SciPyPython} to perform the fitting. The initial guesses for the parameters $A$, $x_0$, and $y_0$ are obtained by finding the brightest pixel within a 1.5\,arcsec (10 pixel) cutout around the coordinates of the UV centroids (Sec.~\ref{sec:uv_cent}). The initial guesses for $\sigma_x$, $\sigma_y$ and $\theta$ are the same for all images, at 2\,px, 2\,px, and $0^{\circ}$ respectively. We then let \textsc{curve\_fit} fit a 2D Gaussian to the masked image starting with the above initial parameters. If the fit converges, the fit centroid is defined as the centre of the 2D Gaussian, ($x_0$, $y_0$).

We find fit uncertainties using a bootstrapping method with 100 trials per galaxy. We first add random Gaussian (with mean \(\sim0\) and \(\sigma \sim\) noise level in the image) noise to each pixel in the input image for each trial, and create 100 ``noisy images'' per galaxy. For each noisy image, we repeat the 2D Gaussian fitting procedure described above, including the noise estimation, \(2\sigma\) masking, initial guess, and Gaussian fit. We exclude trials where the fit fails to converge (which happens for <3 trials out of 100 in our analysis). Then the average centre position over all converged trials gives the final centroid position of the galaxy. The standard deviation among converged trials gives the \(1\sigma\) fit uncertainty on the $x$ and $y$ positions of the centroid. The fit error is then obtained as the sum in quadrature of the $x$ and $y$ uncertainty.

For the continuum emission of DEIMOS\_COSMOS\_881725 and VUDS\_ECDFS\_530029038, we perform the fitting in a smaller crop extent of 2.25\,arcsec (15 pixels) on a side to avoid nearby bright sources.

To test the robustness of our centroid estimate, we also fit a 2D Sersic to the images, using a Sersic model instead of a Gaussian in our fitting code above. The centroids thus obtained agree with our 2D Gaussian centroids to within one pixel, i.e., less than a kpc. We also vary the masking criterion from 1$\sigma$ to 3$\sigma$. The centroids again agree to within $\sim$1.5\,kpc. We choose the 2$\sigma$ masking for our final fits to avoid being unduly influenced by spurious noise features, while at the same time not biasing the result towards the peak pixel.

As a sanity check, we compare the brightest source pixel in each image (i.e., the ``peak'' pixel supplied as the initial guess on the $x$ and $y$ coordinates to the centroid fitting code) and the centre of light \citep[\textsc{scipy.ndimage.center\_of\_mass},][]{Virtanen2020SciPyPython} with the centroid obtained from the fit. We find that the peak pixel and centre of light generally trace the fit centroid position. The peak is found to be within 2\,kpc of the 2D Gaussian (and 2D Sersic) centroids. The centre of light is more susceptible to influence by nearby sources and noise fluctuations, with a large scatter (less than 1\,kpc for high SNR images, and up to $\sim$5.5\,kpc for images with bright nearby sources and strong noise features), but generally following the 2D Gaussian centroid positions.

\subsubsection{Optical}
For optical images, we use the same procedure as for the \cii and FIR continuum to find centroids and uncertainties, with the exception of the crop extent. We use various crop extents between 2.25 and 4.50\,arcsec (15 and 30 pixels) to perform the optical centroid fits, so as to avoid other bright sources close to the target galaxy. Despite these measures, several galaxies (vuds\_cosmos\_510596653, vuds\_cosmos\_5101288969, DEIMOS\_COSMOS\_357722, DEIMOS\_COSMOS\_722679, and DEIMOS\_COSMOS\_843045) fail to fit or return a poor fit (either due to high noise in the image or the presence of bright sources very close to the target galaxy). We therefore exclude the optical emission of these galaxies from further analysis.

\begin{figure*}
     \centering
     \begin{subfigure}[b]{0.32\textwidth}
         \centering
         \includegraphics[width=\textwidth]{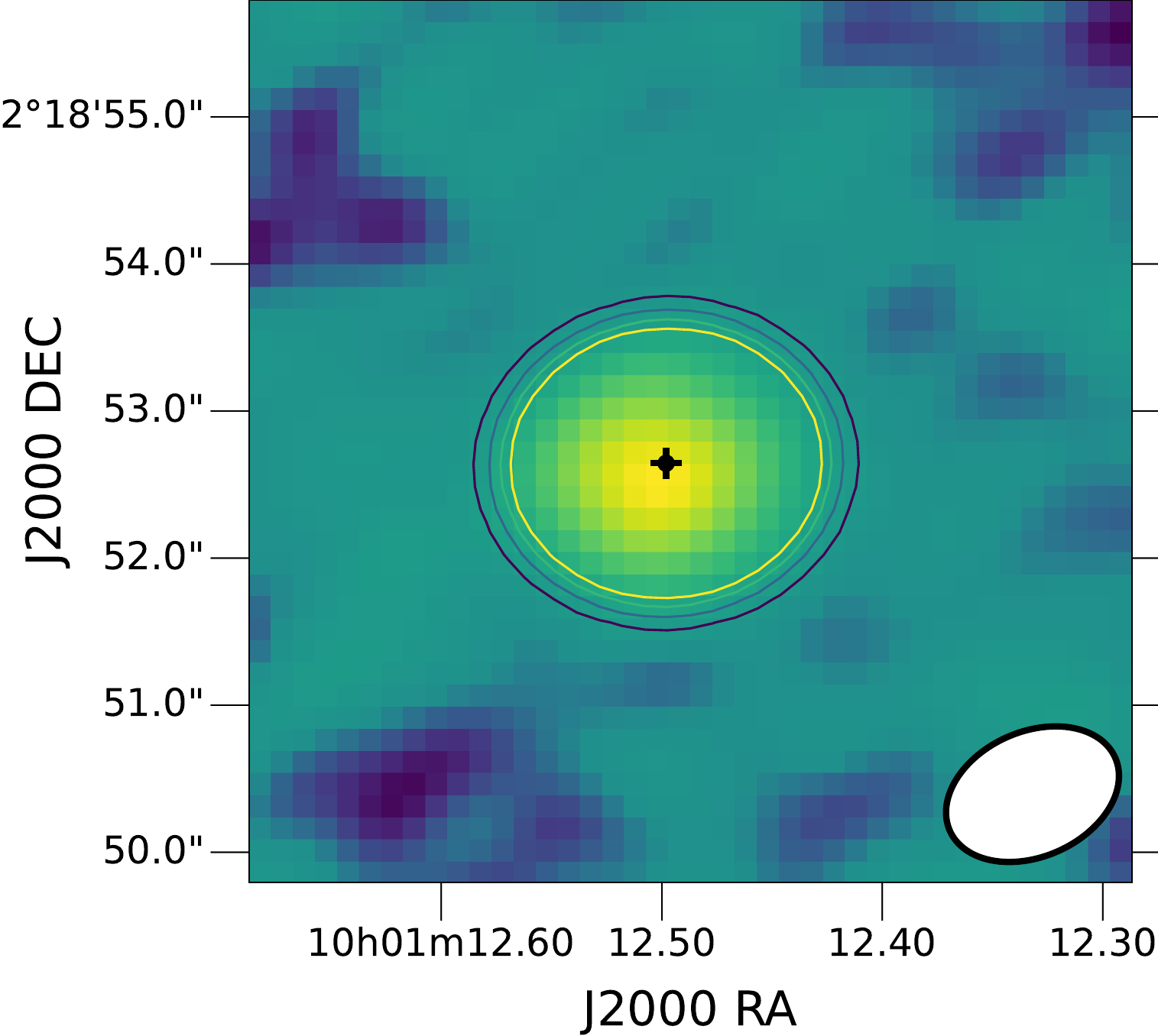}
         \caption{\cii fit}
         \label{fig:cii_fit}
     \end{subfigure}
     \hfill
     \begin{subfigure}[b]{0.32\textwidth}
         \centering
         \includegraphics[width=\textwidth]{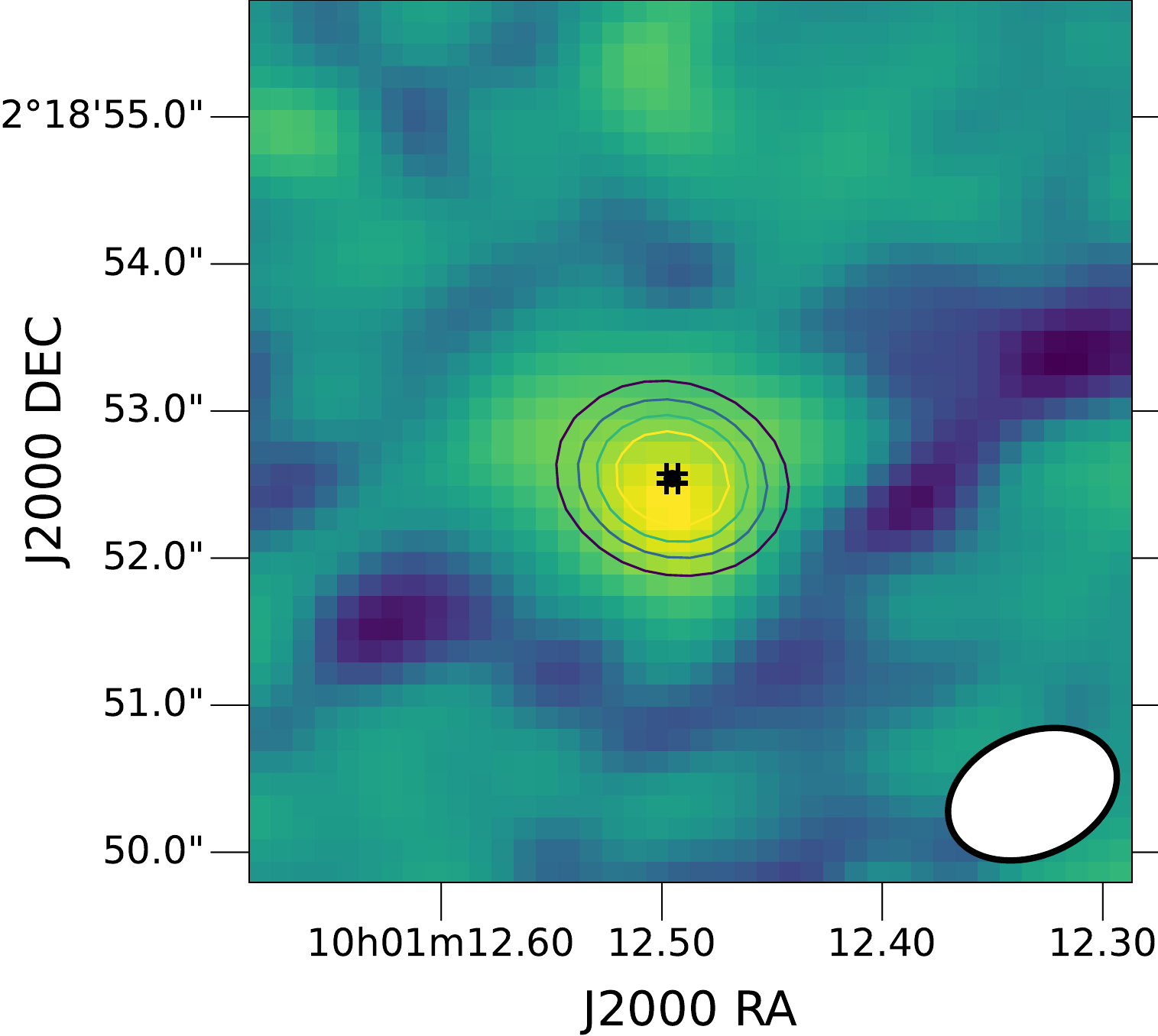}{}
         \caption{FIR continuum fit}
         \label{fig:cont_fit}
     \end{subfigure}
     \hfill
     \begin{subfigure}[b]{0.32\textwidth}
         \centering
         \includegraphics[width=\textwidth]{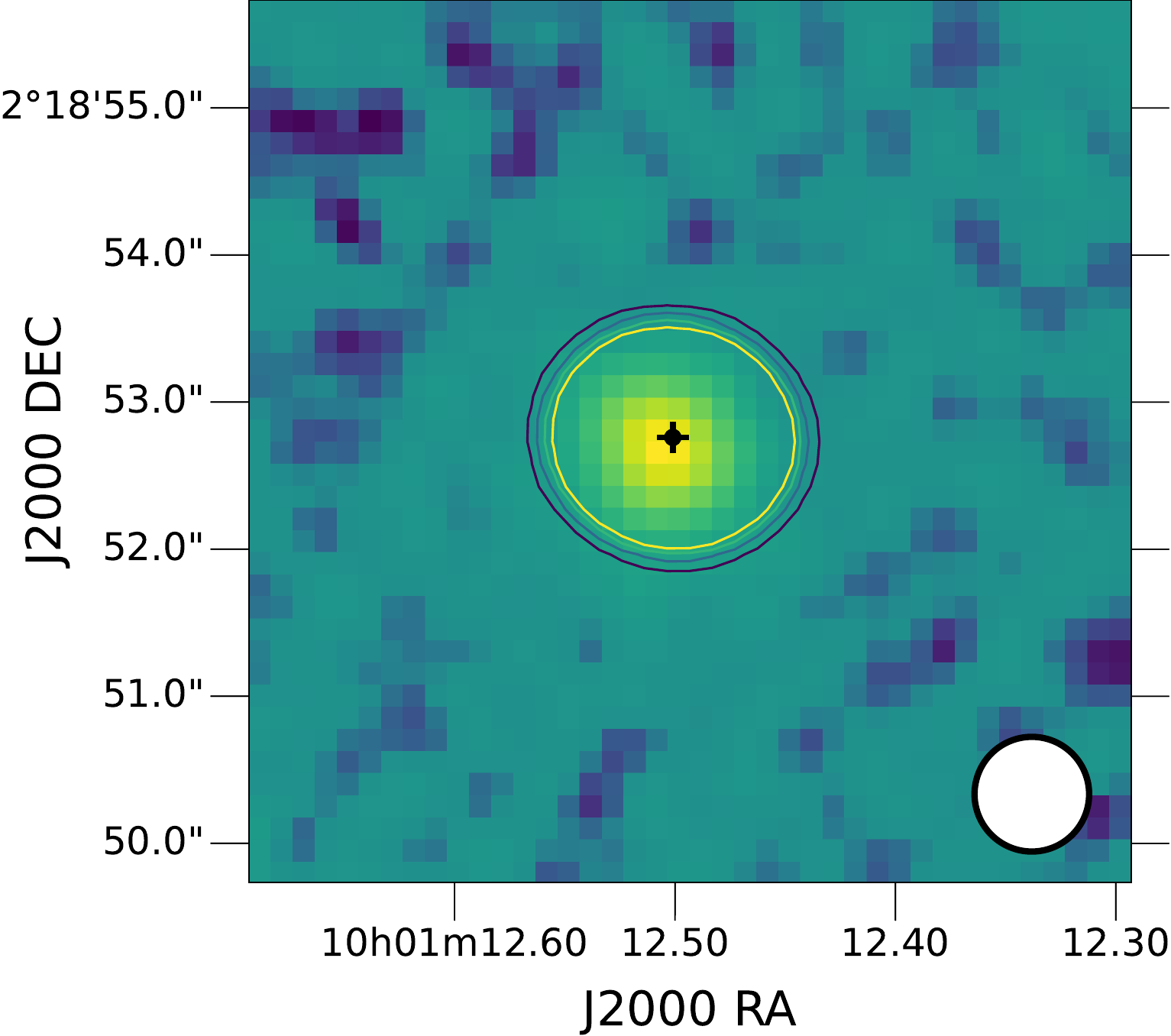}
         \caption{Optical fit}
         \label{fig:opt_fit}
     \end{subfigure}
        \caption{2D Gaussian fits (shown as contours drawn at 2, 3, 4, and 5$\sigma$) to find the centroids of (a) \cii, (b) FIR continuum, and (c) optical emission for the galaxy VUDS\_COSMOS\_5101218326. The fit uncertainty (without positional and $\Delta_{NC}$ uncertainties; see Sec.~\ref{sec:fitting}) on the $x$ and $y$ position of the centroid is also indicated with black errorbars. ALMA and optical beamsizes are shown as filled white ellipses.}
        \label{fig:centroid_fits}
\end{figure*}

In Fig.~\ref{fig:centroid_fits}, we show an example of centroid fits to emission in \cii, FIR continuum, and optical emission for the galaxy VUDS\_COSMOS\_5101218326.

\subsubsection{Positional accuracy}
\label{sec:pos_uncert}
In addition to the fit error, there is positional uncertainty associated with telescope pointings, which contributes to the uncertainty on the centroid position. The positional uncertainty for optical centroids is taken as 80\,mas in both RA and Dec \citep{McCracken2012UltraVISTA:COSMOS}, resulting in a total positional uncertainty of $\sim110$\,mas.

For ALMA images, the positional accuracy $\Delta p$ (in milliarcsec) is given by Eq.~3 of \citet{Faisst2020TheMeasurements}:
\begin{equation}
    \Delta p = \frac{70\,000}{\nu B \sigma_\textit{peak}}
\end{equation} where $\nu$ is the observed frequency in GHz, $B$ is the maximum baseline length in kilometers, and $\sigma_\textit{peak}$ is the calibrator SNR at the peak of the emission. For the ALPINE data, $B$ = 0.2\,km corresponding to the C43-1 configuration, $\nu$ is $\sim 330$\,GHz, and $\sigma_\textit{peak}$ is $\sim50$, which results in a $\Delta p$ of $\sim20$\,mas. We use this value as the positional uncertainty on the \cii and FIR continuum centroids.

\subsubsection{Noise correlation at the beam scale}
\label{sec:noise_correl}
For ALMA images, noise is correlated on the scale of the beam (which is the same size as most of our marginally resolved objects), which introduces additional uncertainty in determining the centroid position. We estimate this uncertainty in the following way. We first inject several artificial 2D elliptical Gaussian galaxies of the same form as in Eq.~\ref{eq:elliptical_gaussian} into each \cii and FIR continuum image around the central source. The 2D Gaussian height, widths in x and y, and position angle of the injected sources are chosen to be random values within 1$\sigma$ of the corresponding properties of the real source in the image as recovered by our fitting in Sec.~\ref{sec:cii_fitting}.
%The size of each of these simulated galaxies is between $\sim$1 and 3.5 times the ALMA beam} In other words, 
%\begin{align}
%    \sigma_x = k \times \frac{b_{\texttt{maj}}}{2.355}
%\end{align}
%and
%\begin{align}
%    \sigma_y = k \times \frac{b_{\texttt{min}}}{2.355},
%\end{align}
%where 
%$b_{\texttt{maj}}$ and $b_{\texttt{min}}$ are the full widths at half-maximum (FWHM) of the major and minor axes of the ALMA beam and \tcg{$k$ is a scaling factor such that $\frac{k}{2.355}$ is a random number between 0.5 and 1.5.}
The centre positions ($x_0$, $y_0$) of the simulated galaxies are chosen at random within an annulus of 6 to 12\,arcsec (40 to 80 pixels), around the centre of the image. We ensure that no two simulated galaxies are within five standard deviations of each other so that the flux from one does not influence the fit of another.

Then, we fit each of these simulated galaxies in the same way as we fit the real galaxy (as described in Sec.~\ref{sec:cii_fitting}) within a crop extent of 3.0\,arcsec (20 pixels). We introduce 20 simulated sources and average over 50 trials per source (a total of 1000 fits per image). We calculate the difference between injected and fit centroid position for each simulated galaxy, and then the sigma clipped median of these differences for all the simulated galaxies in each image. This value is taken as the noise correlation uncertainty ($\Delta_{NC}$) for \cii and FIR continuum centroids. Thus, $\Delta_{NC}$ is calculated individually for each ALPINE source, ensuring that the simulated sources reflect the Gaussian properties of the real source.

% \begin{figure*}
%     \centering
%     \includegraphics[width=\textwidth]{DEIMOS_COSMOS_873756_cii_correlation_fits.pdf}
%     \caption{The ALMA \cii image of the galaxy DEIMOS\_COSMOS\_873756 with several injected simulated galaxies (numbered) is shown in the first panel. The obtained centroid fits for each injected galaxy are shown in the remaining panels labelled with the corresponding number. The injected galaxy positions are shown in white, while the fit positions are shown in black (with fit errorbars). The white double headed arrows at the top right indicate 1 arcsec, shown for scale. The ALMA beam is shown in the first panel as a filled white ellipse.}
%     \label{fig:simulated}
% \end{figure*}

\begin{figure}
    \centering  
    \includegraphics[width=\columnwidth]{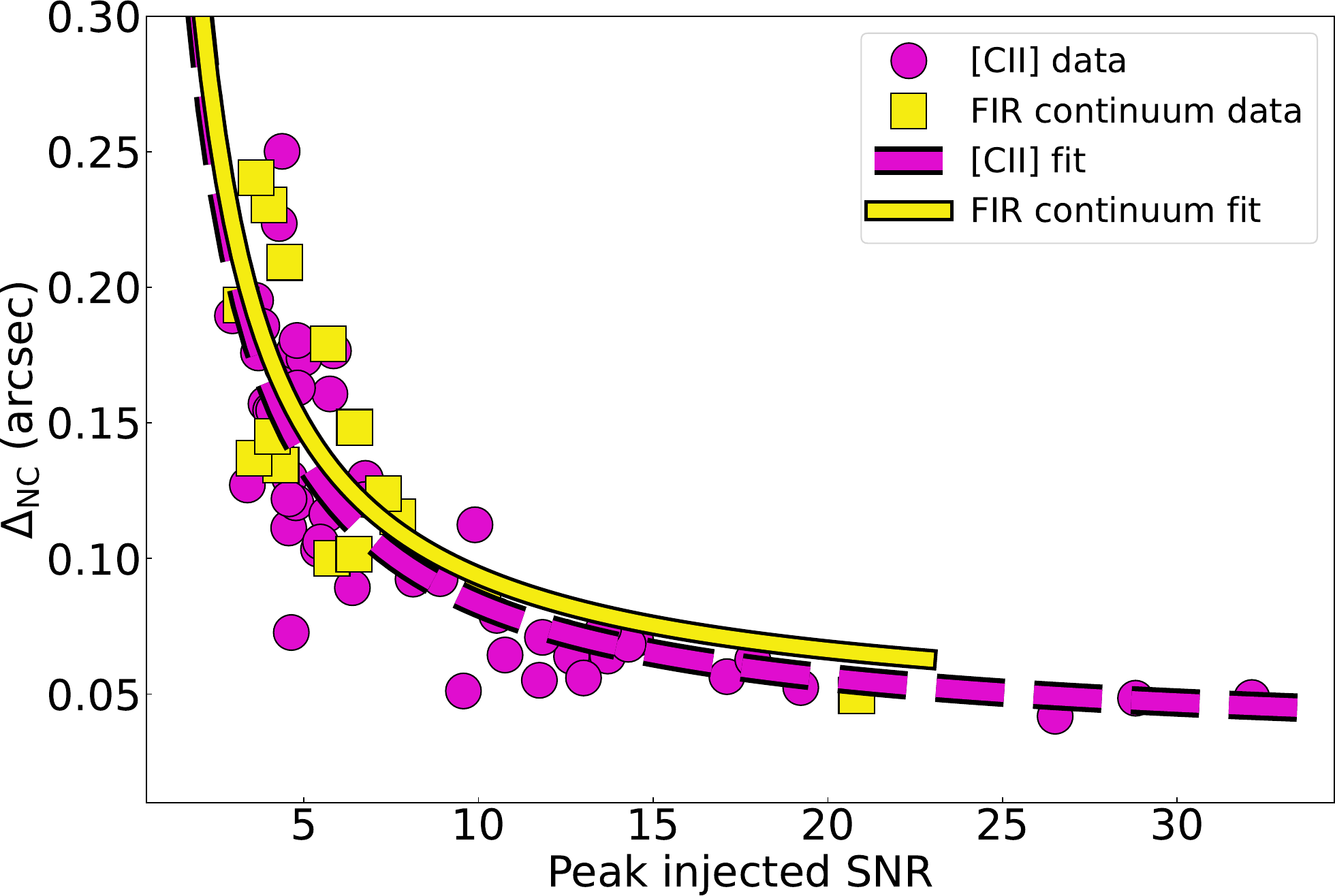}
    \caption{The inverse relationship between the noise correlation uncertainty ($\Delta_{NC}$; see Sec.~\ref{sec:noise_correl}) and median peak SNR of the injected Gaussians. The results for \cii and FIR continuum images are shown in pink and yellow.}
    \label{fig:noise_correl_snr}
\end{figure}

% In Fig.~\ref{fig:simulated}, we show an example of an image with several simulated galaxies and the centroid fits for each.
In Fig.~\ref{fig:noise_correl_snr}, we plot $\Delta_{NC}$ calculated in each \cii and FIR continuum image against the median peak SNR (i.e., height) of the injected Gaussians in that  image. The plot follows an inverse relation (of the form $a + \frac{b}{x}$, where $x$ is the peak SNR, and $a$ and $b$ are constants), where the images with the lowest SNR show the largest deviation between injected and fit centroid positions. This is expected because given that the sizes of noise peaks and troughs are comparable to the ALMA beam, the morphology of a low SNR source will be more easily perturbed by the noise, resulting in a larger positional offset in the fit. Hence, the probability of positional offset of a fit has an anti-correlation with SNR.

\subsubsection{Total uncertainty}
\label{sec:tot_uncert}
The total uncertainty on the optical centroids is calculated as the sum in quadrature of the bootstrapped fit uncertainty (Sec.~\ref{sec:fitting}) and the positional uncertainty (Sec.~\ref{sec:pos_uncert}). For the ALMA \cii and FIR continuum centroids, the noise correlation term (Sec.~\ref{sec:noise_correl}) is also added in quadrature. In general, the noise correlation uncertainty dominates over the positional and fit uncertainties, especially for low SNR sources.

\subsection{Spatial offsets}
The spatial offset between any two emissions is defined as the spatial separation (in arcsec) between the calculated centroid positions of the two emissions. We use the {\sc astropy} \citep{TheAstropyCollaboration2022ThePackage} function {\sc coordinates.SkyCoord.separation} to estimate this separation. The corresponding uncertainty on the offset is calculated as the sum in quadrature of the total positional uncertainties (see Sec.~\ref{sec:tot_uncert}) on the centroids of the two emissions.

\section{Results}
% \subsection{Identifying galaxies with significant offsets}
\label{sec:significant_offsets}
Given a Gaussian uncertainty $\sigma$ on each coordinate (RA and DEC), the expected distribution of offsets (r) is given by a 2D circular Gaussian of the form:

\begin{align}
    f(r)\ dr = 2 \pi r\ \left(\frac{1}{2 \pi \sigma^2}\right)\ e^{\left(-\frac{1}{2} \left(\frac{r}{\sigma}\right)^2\right)}\ dr\\
    % = \left(\frac{r}{\sigma^2}\right) e^{\left(-\frac{1}{2} \left(\frac{r}{\sigma}\right)^2\right)} dr\\
    = \left(\frac{r}{\sigma}\right)\ e^{\left(-\frac{1}{2} \left(\frac{r}{\sigma}\right)^2\right)}\ d\left(\frac{r}{\sigma}\right)
\end{align}
Calculating the significance ($s$) as measured offset divided by the measured total uncertainty, i.e., $s = \frac{r}{\sigma}$, we have,

\begin{align}
\label{eq:circ}
    f(r)\ dr = s\ e^{\left(-\frac{1}{2} s^2\right)}\ ds.
\end{align}

As the uncertainties for the various galaxies are not the same, we create a distribution in which each offset is normalised to its $\sigma$. We thus obtain an expected distribution of normalised offsets to compare with our observations (see Fig.~\ref{fig:histograms}). We adopt a \(3\sigma\) threshold to identify offsets that are very likely to be real. In the following analysis, we will call these ``significant'' offsets. Based on the above expected distribution, the fraction of galaxies with no real offset having an observed significance of $s > 3$ should be 0.01. This corresponds to $0.01 \times$ 54 (which is the total number of galaxies in our sample; see Sec.~\ref{sec:sample}) $\sim 0.5$, i.e., less than 1 galaxy. Therefore, we can be confident that all the galaxies in our sample with $s > 3$ have significant offsets unlikely to occur by chance. It is important to note that a significant offset is not necessarily a \emph{large} offset, but it is significant compared to the uncertainty. In other words, the sensitivity and accuracy of our analysis increases with SNR.

\subsection{Offset distributions}
\begin{figure*}
    \centering
    \includegraphics[width=\textwidth]{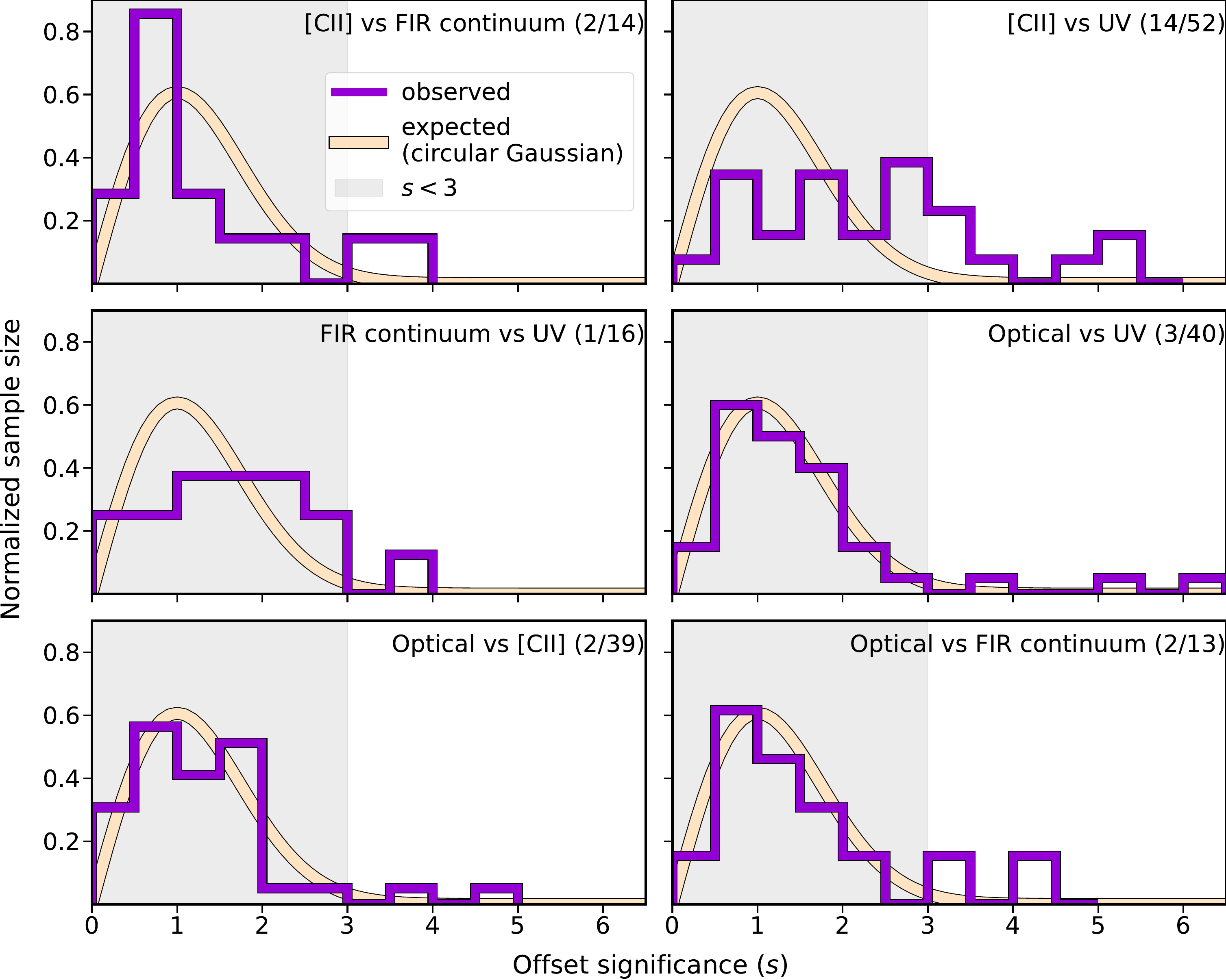}
    \caption{Normalised histograms of the significance ($s$) of observed spatial offsets between (from left to right; top) \cii-FIR continuum, \cii-UV, (middle) FIR continuum-UV, Optical-UV, (bottom) Optical-\cii, and Optical-FIR continuum emission are shown in violet. The cream-coloured curve is the expected distribution of offsets, modelled with a circular Gaussian (as described in Sec.~\ref{sec:significant_offsets})). Although the majority of galaxies lie within 3\(\sigma\) of the expected distribution (shown by the grey-shaded region), there still exists a tail of galaxies with significant offsets ($s>3$). The number of galaxies with significant offsets is given as a fraction of total number of galaxies with that particular offset measurement. For instance, there are 14 galaxies in our sample with a measurement of \cii-FIR continuum offset, of which two are significant, written as ``\cii vs FIR continuum (2/14)'' in the top left panel.}
    \label{fig:histograms}
\end{figure*}

In Fig.~\ref{fig:histograms}, we show histograms of the significance of spatial offsets between combinations of \cii, UV, Optical, and FIR continuum. We plot the expected distribution as a normalised circular Gaussian (Eq.~\ref{eq:circ}) and indicate our \(3\sigma\) threshold using a grey shaded region. We find that for the majority of galaxies, the observed offsets could be caused by measurement uncertainties. However, some galaxies lie outside the expected distribution with a significance of \(s > 3\).

% \begin{figure*}
%      \centering
%      \begin{subfigure}[b]{0.49\textwidth}
%          \centering
%          \includegraphics[width=\textwidth]{cii_uv_cent_tail.pdf}
%          \caption{\cii-UV}
%          \label{fig:cii_uv_hist}
%      \end{subfigure}
%      \hfill
%      \begin{subfigure}[b]{0.49\textwidth}
%          \centering
%          \includegraphics[width=\textwidth]{cii_cont_cent_tail.pdf}
%          \caption{\cii-FIR continuum}
%          \label{fig:cii_cont_hist}
%      \end{subfigure}
%      \medskip
%      \begin{subfigure}[b]{0.49\textwidth}
%          \centering
%          \includegraphics[width=\textwidth]{cont_uv_cent_tail.pdf}
%          \caption{FIR continuum-UV}
%          \label{fig:cont_uv_hist}
%      \end{subfigure}
%      \hfill
%      \begin{subfigure}[b]{0.49\textwidth}
%          \centering
%          \includegraphics[width=\textwidth]{opt_uv_cent_tail.pdf}
%          \caption{Optical-UV}
%          \label{fig:opt_uv_hist}
%      \end{subfigure}
%      \medskip
%      \begin{subfigure}[b]{0.49\textwidth}
%          \centering
%          \includegraphics[width=\textwidth]{opt_cont_cent_tail.pdf}
%          \caption{Optical-FIR continuum}
%          \label{fig:opt_cont_hist}
%      \end{subfigure}
%      \hfill
%      \begin{subfigure}[b]{0.49\textwidth}
%          \centering
%          \includegraphics[width=\textwidth]{opt_cii_cent_tail.pdf}
%          \caption{Optical-\cii}
%          \label{fig:opt_cii_hist}
%      \end{subfigure}

\begin{figure*}
     \centering
     \begin{subfigure}[b]{0.49\textwidth}
         \centering
         \includegraphics[width=\textwidth]{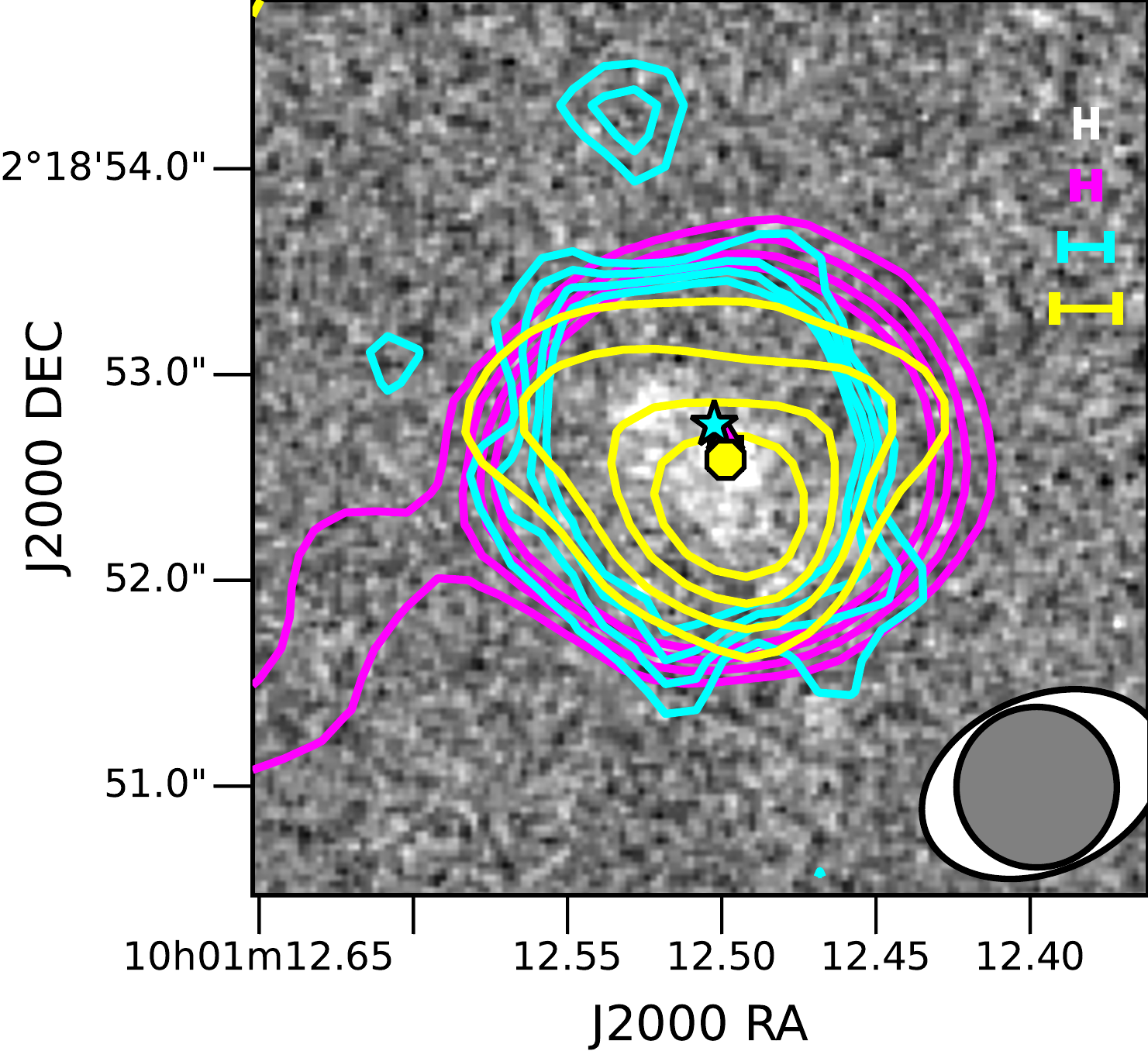}
         \caption{No significant offset}
         \label{fig:no_offset}
     \end{subfigure}
     \hfill
     \begin{subfigure}[b]{0.49\textwidth}
         \centering
         \includegraphics[width=\textwidth]{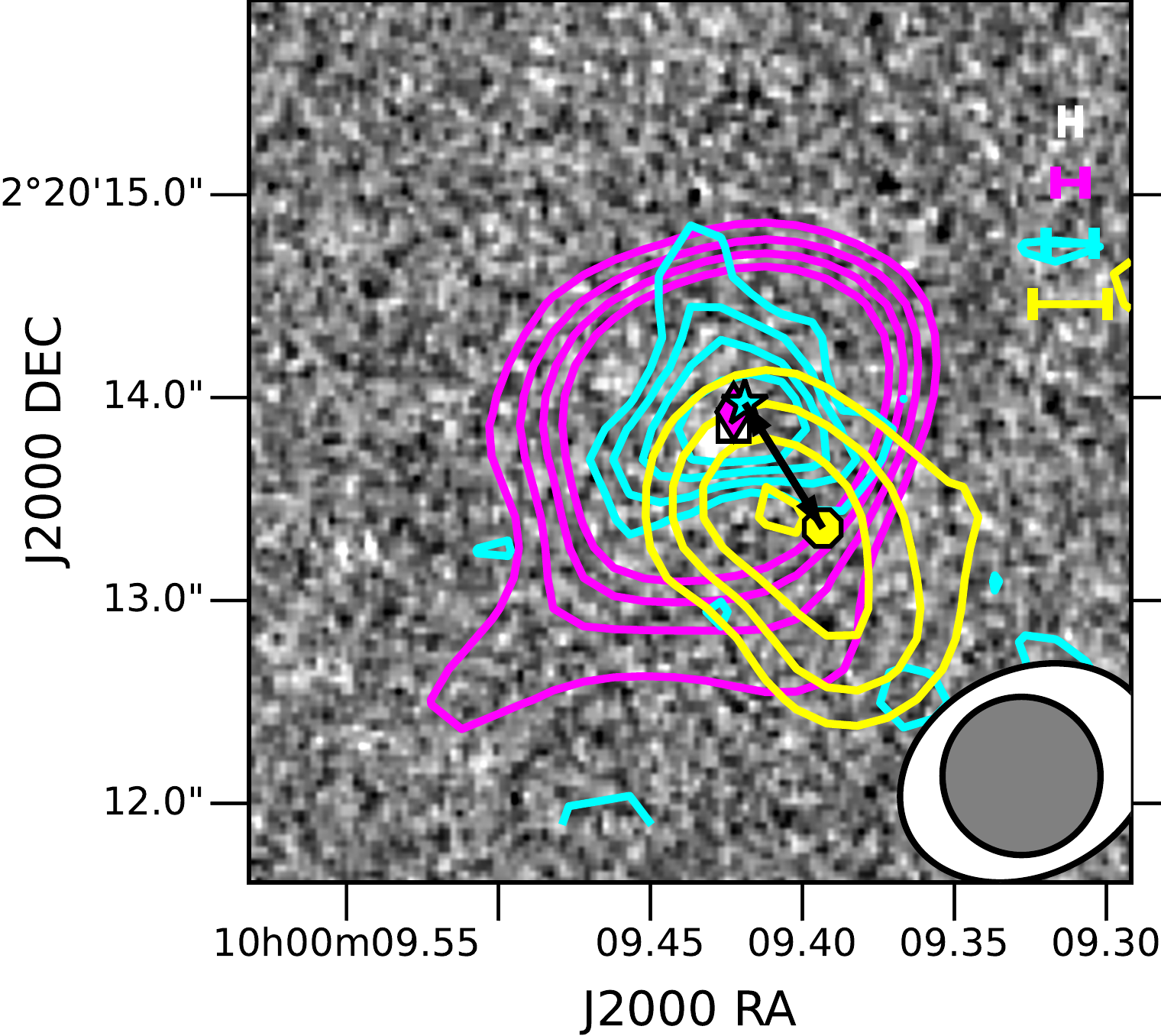}
         \caption{Significant offset}
         \label{fig:sig_offset}
     \end{subfigure}
\caption{(a) A galaxy (VUDS\_COSMOS\_5101218326) with no significant offset among any of the emissions (see Sec.~\ref{sec:significant_offsets}) vs (b) a galaxy (DEIMOS\_COSMOS\_683613) with a significant offset between FIR continuum and all other emissions. UV \emph{HST} image is shown as a grey-scale background with \cii (fuchsia), optical (cyan), and FIR continuum (yellow) overlaid. The contours are drawn at 2, 3, 4, and 5 times the standard deviation (calculated in an annulus with radii of 4.5 and 9.0\,arcsec (30 and 60 pixels), around the centre of the image). The centroids are marked with a white square for UV emission, fuchsia diamond for \cii, cyan star for optical, and yellow octagon for FIR continuum (same colours as the respective contours), and any significant spatial offset among them is indicated by a black double-headed arrow. The calculated total error (see Sec.~\ref{sec:tot_uncert}) in each emission is indicated on the top right in the same colour as the centroids. The ALMA (\cii and FIR continuum) beam is shown as a filled white ellipse, and optical beam as a filled grey circle.}
\label{fig:overlaid_contours}
\end{figure*}

Fig.~\ref{fig:overlaid_contours} shows an example of one galaxy (VUDS\_COSMOS\_5101218326) with no significant offset ($s < 3$) and one (DEIMOS\_COSMOS\_683613) with a significant offset ($s > 3$). We see that one has all centroids close together (offsets \(\lesssim\) 1.2\,kpc), while the other has FIR continuum centroid separated from the other centroids (offset \(\sim\)4\,kpc). As the FIR continuum traces dust, it appears that this second galaxy (DEIMOS\_COSMOS\_683613) has the bulk of its dust offset from stars and gas (both atomic and ionised gas as traced by \cii). Several other significant offset galaxies are shown in Fig.~\ref{fig:overlaid_contours2} of Appendix~\ref{apx:significant_offsets}.

In Table~\ref{tab:subsamples_offs}, we show the number of galaxies in our sample that display such significant spatial offsets between pairs of emission tracers. We also specify the median offset and uncertainty for the full distributions of offsets, and the median of only the significant offsets in each distribution. We find the largest number of significant offsets between \cii-UV, followed by UV-optical. This may be because the uncertainties on centroids are smaller in UV and optical, and our sample size is largest in the UV, \cii, and optical. Overall, \(\sim\)30 percent of the galaxies in our sample (17 galaxies) show significant offsets between at least two emissions, while the remaining \(\sim\)70 percent (37 galaxies) do not have significant offsets between any two emissions.

\begin{table*}
\centering
% \caption{Offset Distribution}
    \begin{tabular}{c|c|c|c|c|c}
    \hline
    Emission tracer pair & Median offset & Median uncertainty & Number of $s>3$ galaxies/ & \multicolumn{2}{c}{Median significant offset}\\
    & (arcsec) & (arcsec) & Total number of galaxies & (arcsec) & (kpc) \\
    & & & (percentage) & & \\
    \hline
    \cii-FIR continuum & 0.13 & 0.17 & 2/14 (14 percent) & 0.56 & 3.53 \\
    \cii-UV & 0.22 & 0.13 & 14/52 (27 percent) & 0.54 & 3.63 \\
    FIR continuum-UV & 0.24 & 0.16 & 1/16 (6 percent) & 0.67 & 4.09 \\
    Optical-UV & 0.15 & 0.13 & 3/40 (8 percent) & 0.67 & 4.47 \\
    Optical-\cii & 0.17 & 0.17 & 2/39 (5 percent) & 0.72 & 4.56 \\
    Optical-FIR continuum & 0.3 & 0.19 & 2/13 (15 percent) & 0.64 & 4.09 \\
    \end{tabular}
    \caption{Number of galaxies with significant spatial offsets. The first column gives the two emission tracers between which we calculate the offset. Second and third columns give the sigma-clipped medians of the full distribution of offsets and uncertainties (Sec.~\ref{sec:fitting}) respectively. Fourth column gives the number of galaxies that show significant offsets (see Sec.~\ref{sec:significant_offsets}) out of the number of galaxies for which we calculate this offset, with the percentage specified in parentheses. The last column gives the sigma-clipped median of only the significant offsets both in arcsec and kpc.}
    \label{tab:subsamples_offs}
\end{table*}
% 2.1951719, 4.3903439, 3.3241175, 2.2578911, 4.0767479, 3.7631519 kpc

\section{Discussion}
\subsection{Relating spatial offsets to physical properties}
\label{sec:phys}
We now consider the physical origin of spatial offsets in distant galaxies. Wherever possible, we plot galaxy physical properties such as specific star-formation rate (sSFR), stellar mass ($M_\star$), UV continuum slope ($\beta$), etc. from the \citet{Faisst2020TheMeasurements} and \citet{Bethermin2020TheProperties} catalogues against the measured spatial offsets. We then look for trends in these plots that may reveal the phenomenon that is producing spatial offsets. The galaxy physical properties were derived via SED fitting (see Sec.~\ref{sec:alpine}) where galaxies with offsets were treated no differently than others. % Although the ALMA data were not used in the fitting, if the offsets between short and long wavelengths are severe enough, it might affect the derived properties, which in turn would affect whether we observe correlations between offsets and physical properties. However, the implications that these extreme offsets have on SED fitting analysis and the galaxy properties derived are out of the scope of this paper.
% \subsubsection{Spatial offsets driven by physical phenomena}
In the following sections, we describe several potential phenomena that may be driving the observation of spatial offsets (several of these effects may be related to each other).% We start with the most likely scenarios, and then discuss other, less likely, but still plausible scenarios for completeness.
\begin{figure*}
     \centering
     \begin{subfigure}[b]{0.49\textwidth}
         \centering
         \includegraphics[width=\textwidth]{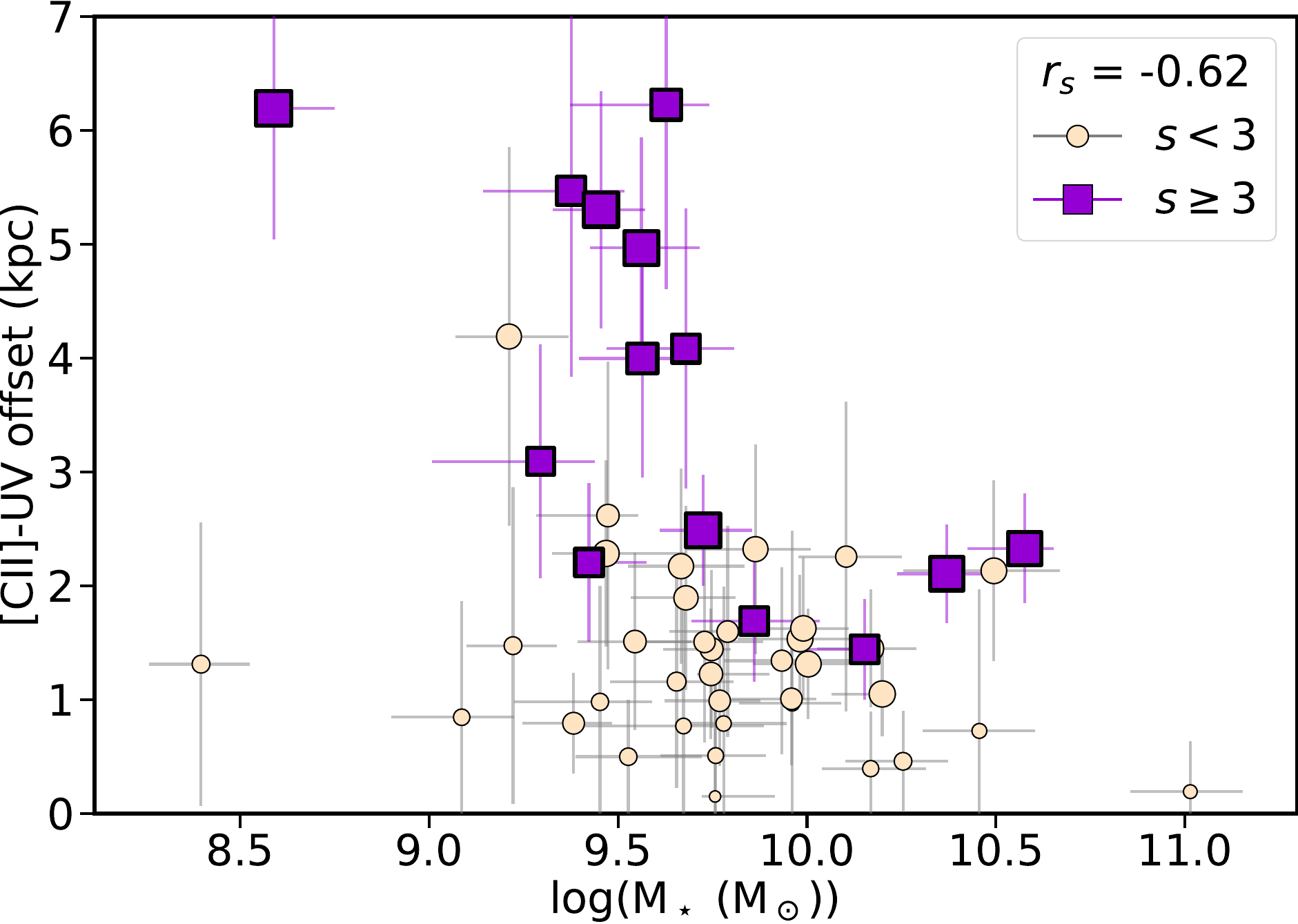}
         \caption{Stellar mass vs \cii-UV offset}
         \label{fig:mstar_cii_uv}
     \end{subfigure}
     \hfill
     \begin{subfigure}[b]{0.49\textwidth} %0.36 is somehow maximum
         \centering
         \includegraphics[width=\textwidth]{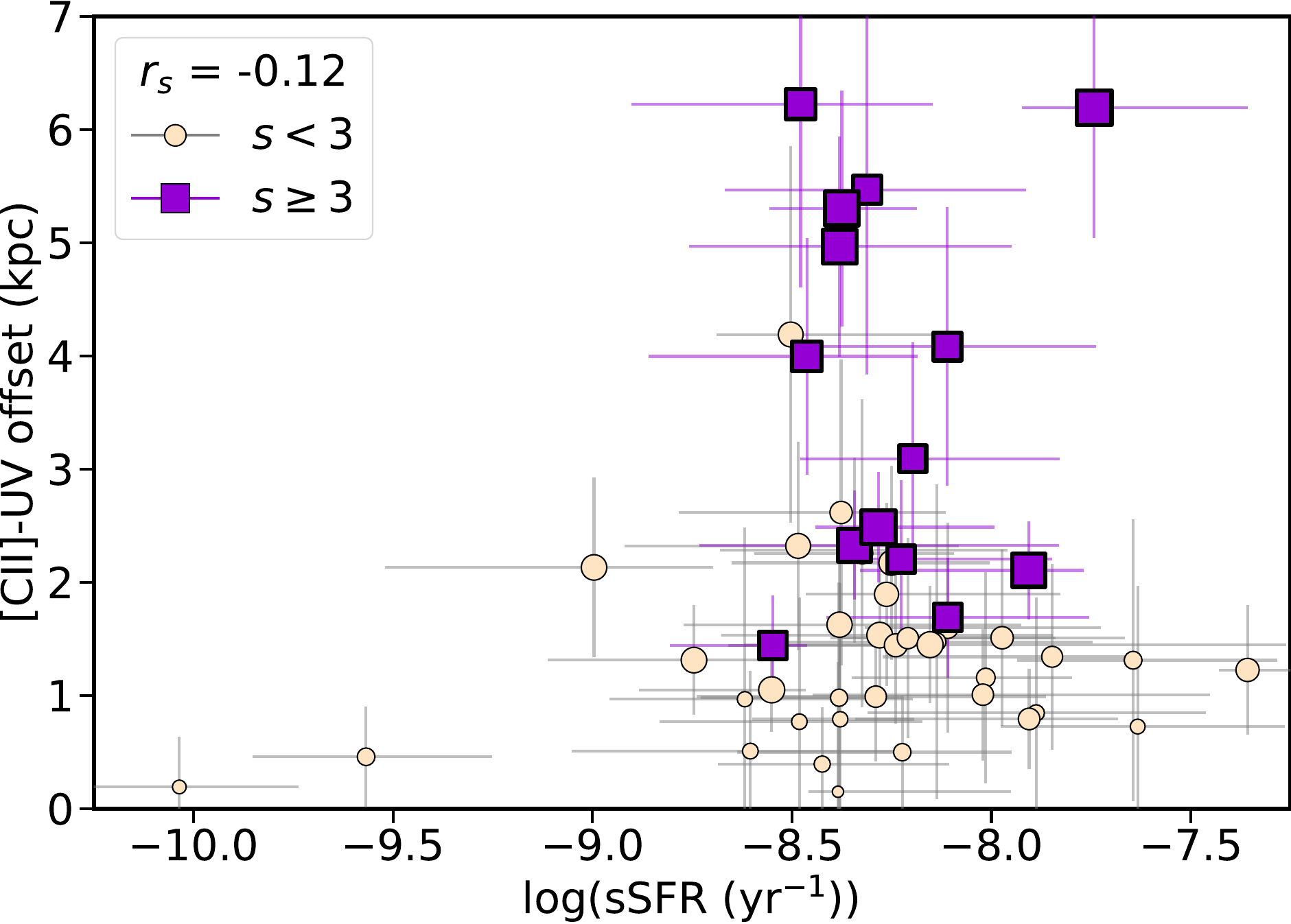}
         \caption{sSFR vs \cii-UV offset}
         \label{fig:ssfr_cii_uv}
     \end{subfigure}
     \medskip
     \begin{subfigure}[b]{0.49\textwidth}
         \centering
         \includegraphics[width=\textwidth]{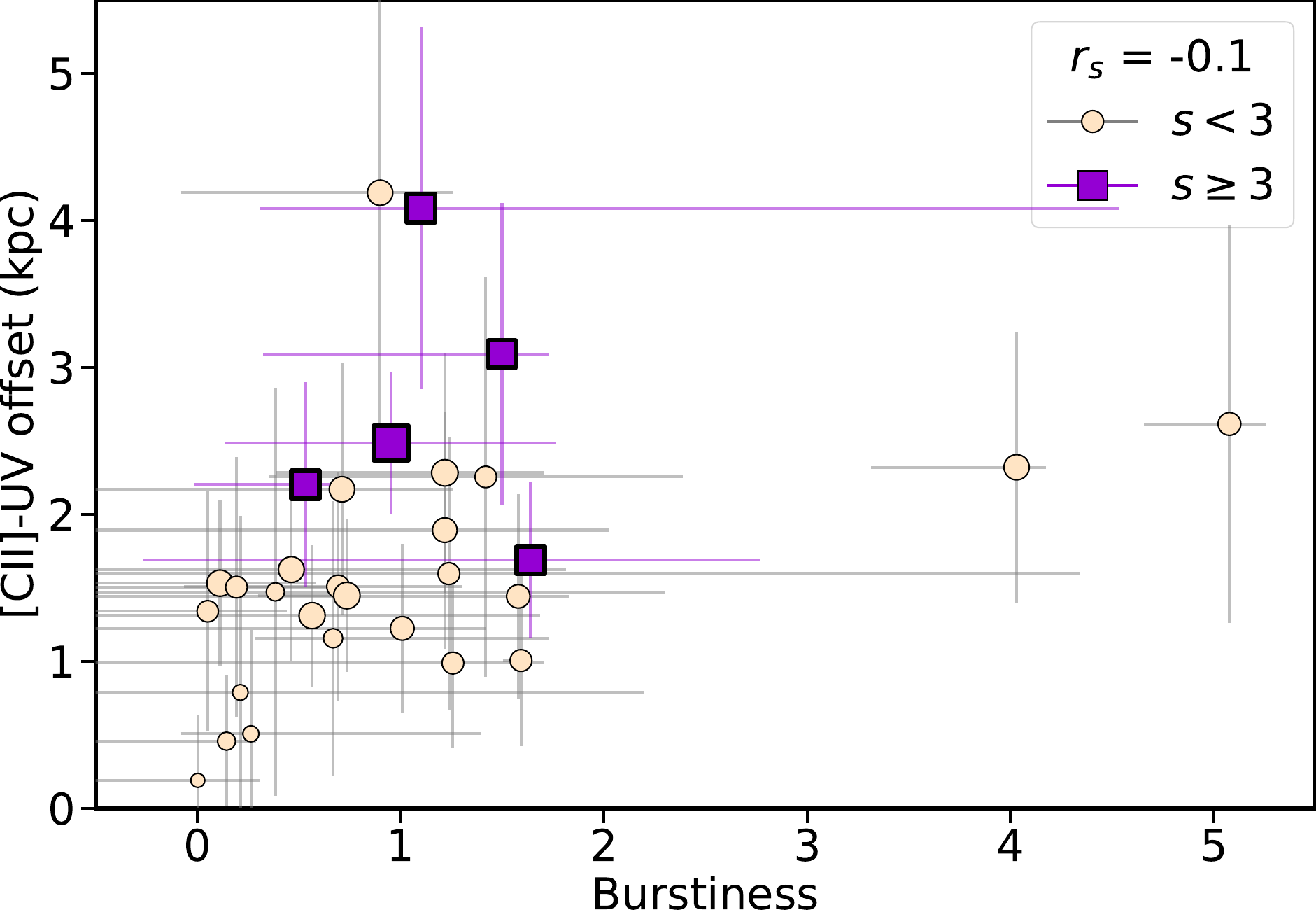}
         \caption{Burstiness vs \cii-UV offset}
         \label{fig:burst_cii_uv}
     \end{subfigure}
     \hfill
     \begin{subfigure}[b]{0.49\textwidth}
         \centering
         \includegraphics[width=\textwidth]{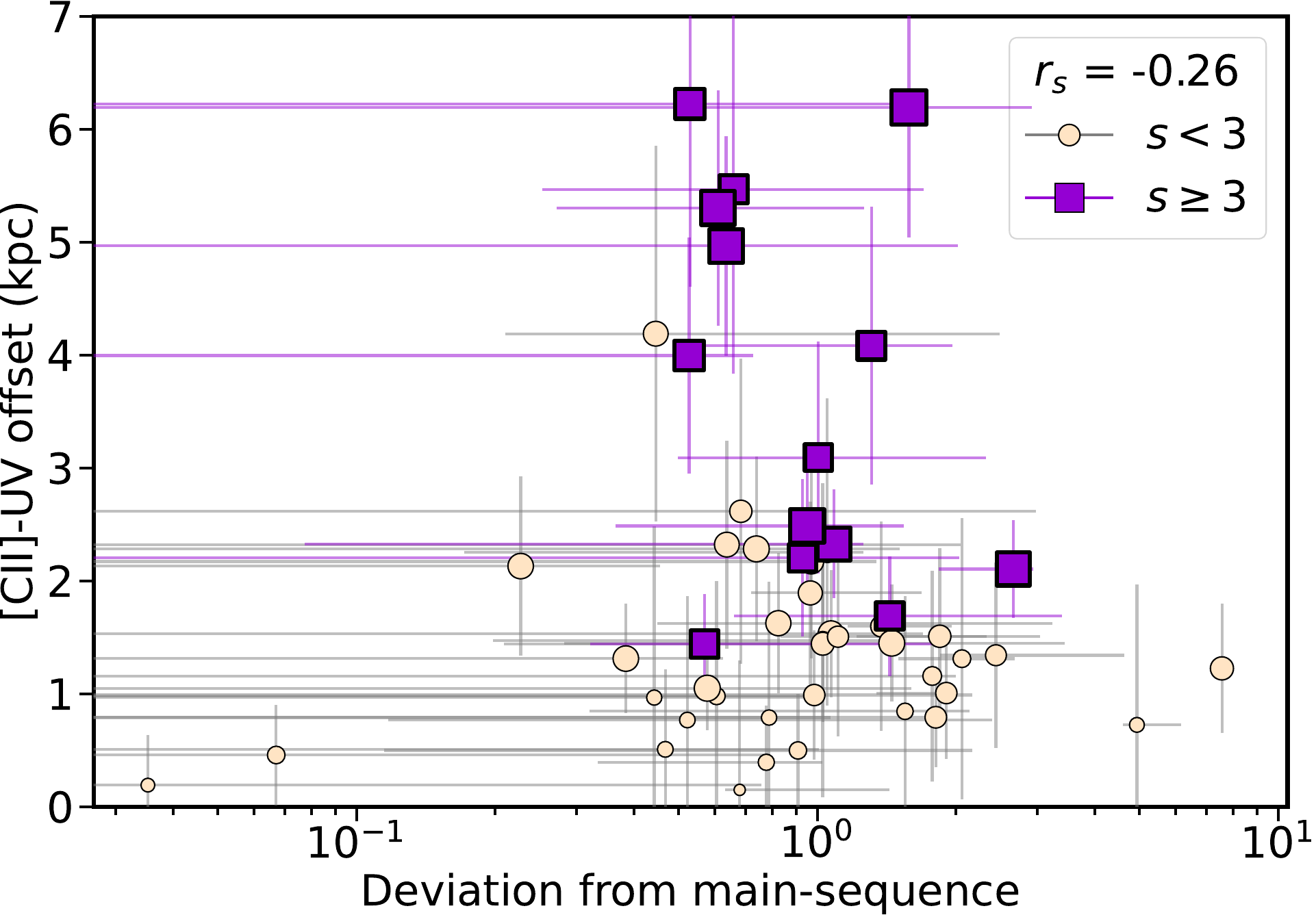}
         \caption{Deviation from main-sequence vs \cii-UV offset}
         \label{fig:devMS_cii_uv}
     \end{subfigure}
     \medskip
     \begin{subfigure}[b]{0.49\textwidth}
         \centering
         \includegraphics[width=\textwidth]{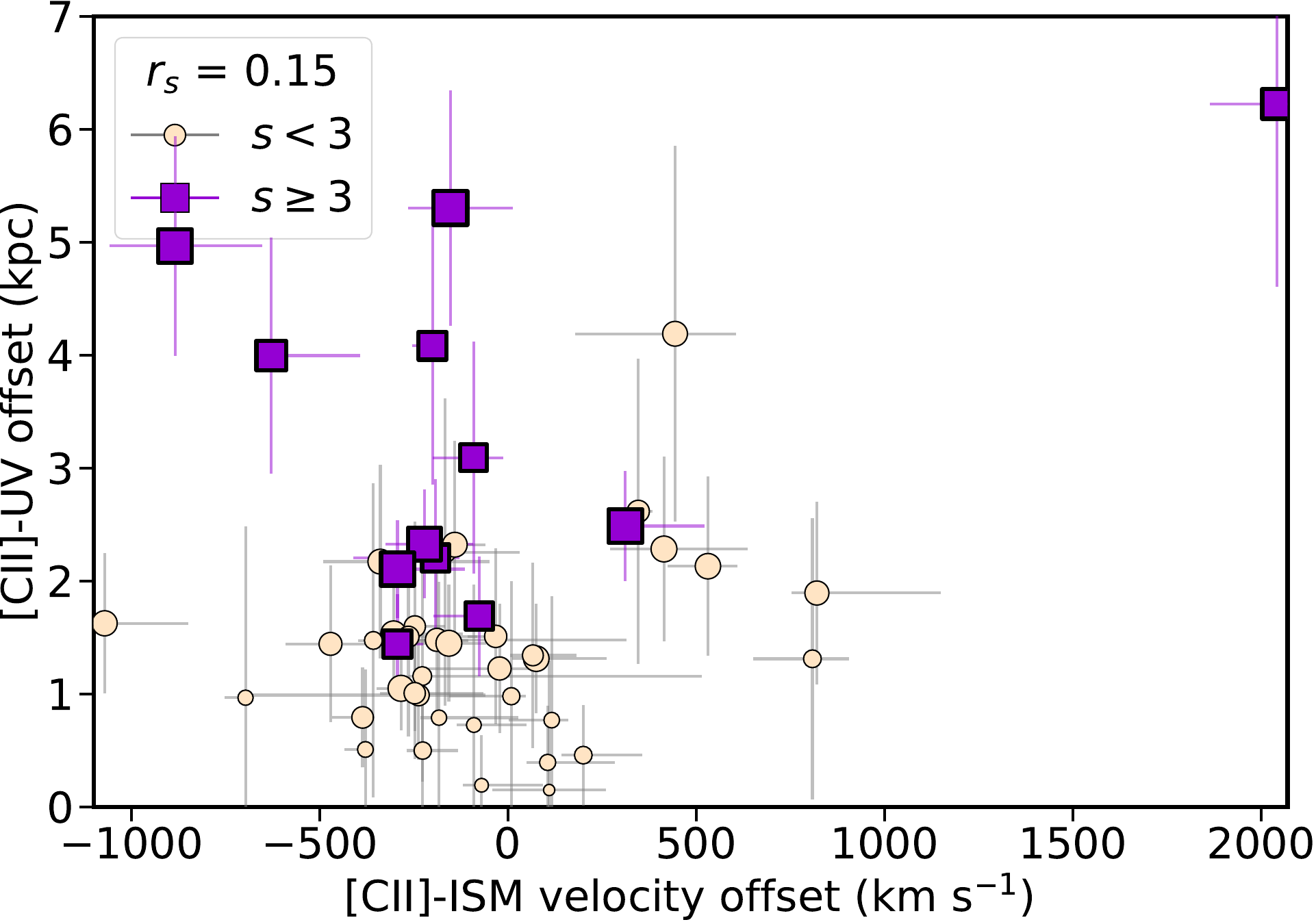}
         \caption{Velocity offset vs \cii-UV offset}
         \label{fig:veloff_cii_uv}
     \end{subfigure}
     \hfill
     \begin{subfigure}[b]{0.49\textwidth}
         \centering
         \includegraphics[width=\textwidth]{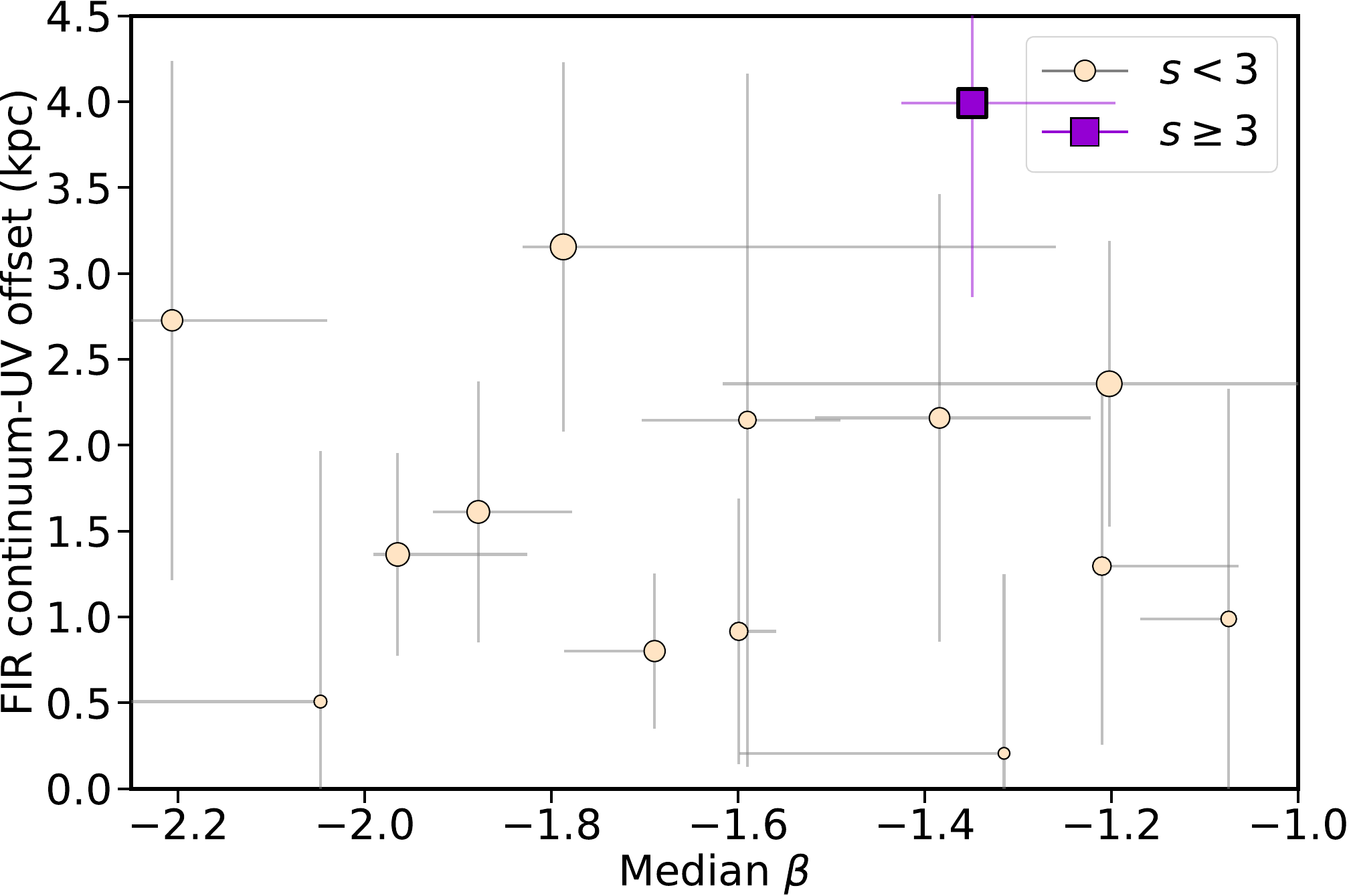}
         \caption{Median $\beta$ vs FIR continuum-UV offset}
         \label{fig:beta_cont_uv}
     \end{subfigure}
\caption{Relating spatial offsets with galaxy physical properties: (a) Stellar mass vs \cii-UV (b) SED-derived sSFR vs \cii-UV (c) burstiness vs \cii-UV (d) deviation from main-sequence vs \cii-UV (e) Velocity offset between ISM absorption lines and \cii vs \cii-UV (f) median $\beta$ vs FIR continuum-UV. The measurements for the galaxies with significant offsets ($s \geq 3$; see Sec.~\ref{sec:significant_offsets}) are shown as violet squares, while galaxies with no significant offsets ($s < 3$) are shown as cream-coloured circles (markersize is proportional to $s$). The uncertainties are plotted as violet and grey errorbars respectively. If there are three or more galaxies with significant offsets, the Spearman's rank coefficient $r_s$ for these is also given.%\tcr{I have added these Spearman rank coefficients based on feedback we received last time, but I believe there are too few significant offset galaxies to really claim any correlations. Should we remove them? If not, ideas on how to interpret these numbers/correlations are more than welcome.}
}
\label{fig:scatter}
\end{figure*}

\subsubsection{Feedback and outflows}

Feedback from star formation, supernovae, or AGN might be physically pushing the enriched gas and/or dust away from the stars \citep{Ceverino2009THEGALAXIES, Maiolino2015TheALMA, Katz2017InterpretingReionization, Vallini2017MolecularEmission, Li2018StellarProperties, Torrey2020TheISM}, which might then produce an observable spatial offset between UV/optical emission compared to the gas (\cii) and dust emission (FIR continuum). Thus, we may expect a large offset in galaxies with high star-formation or AGN activity (assuming that enough time has elapsed since the starburst for the feedback to push the gas/dust to large galactocentric distances). As the galaxies in ALPINE are selected to have low AGN activity \citep{Shen2022The4.45.9}, it is unlikely that the offsets seen here are due to AGN, but high star-formation activity can still clear out gas.% Interestingly, the two ALPINE sources with AGN signatures (FIR continuum detection with no \cii detection \tcm{reference}) are not identified in our analysis as $s > 3$ - \tcr{what does this mean?}.

To test this idea, in Fig.~\ref{fig:mstar_cii_uv} and ~\ref{fig:ssfr_cii_uv}, we plot the stellar mass and total sSFR respectively as functions of the \cii-UV offset. The significant offset sources in Fig.~\ref{fig:mstar_cii_uv} display a Spearman's rank coefficient suggesting an apparent correlation ($r_s = -0.62$), but there does not seem to be a clear distinction between galaxies with and without significant offsets. 
%SFR (calculated assuming a Chabrier IMF, with $L_{\text{IR}}$ taken from ALMA FIR continuum when it is detected, and from \citealt{Fudamoto2020TheZ4.4-5.8} IRX-$\beta$ correction when not detected) % - which prescription are you using to convert LFIR to SFR? e.g., Kennicutt+98, Madau+14 - MR % SFR for SFR of DEIMOS\_COSMOS\_722679 is taken from its SED since $L_{UV}$ is uncalibrated.
We do not observe a clear trend between offset and sSFR. \citet{Carniani2017ExtendedALMA} study the spatial offset in BDF-3299, a Lyman-break galaxy at z=7.1 and compare it to other observations from literature (see their fig. 6). They also do not find any clear correlations between SFR and spatial offset between \cii/\oiii and UV.% - their offsets may not be 'significant' though, so can we compare it to our analysis?}
%note that generally OIII is totally overlapping with the UV bright patches in the ISM of galaxies, so you might compare your CII vs UV offset with their CII vs OIII quite safely

%However, galaxies with very high SFR seem to have relatively small offsets.\tcr{ - comparison with \citet{Carniani2017ExtendedALMA}}.

In Fig.~\ref{fig:burst_cii_uv}, we now plot ``burstiness'' against the \cii-UV offset. The burstiness parameter \citep{Smit2016INFERREDFUNCTIONS, Faisst2019Measurements, Atek2022TheGalaxies} is calculated as a ratio between two SFR indicators: the \ha SFR arising from young stars, and sensitive to star-formation on short-timescales (few Myr), and the UV SFR tracing the stellar continuum, which is sensitive to star-formation on longer timescales (few tens to hundreds of Myr). This parameter therefore measures how instantaneous the star formation is \citep{Kennicutt2012StarGalaxies}, with a ratio above unity suggesting a recent burst, i.e., an episode of star-formation \citep{Atek2022TheGalaxies}. ALPINE galaxies are generally not strongly bursty, and even among those that are, we find no correlation with offset.

Next, in Fig.~\ref{fig:devMS_cii_uv}, we plot the deviation from main-sequence against \cii-UV offset. We compute the deviation as the ratio of the measured SFR vs that expected at the main-sequence, given the stellar mass, and assuming the \citet{Speagle2014AZ0-6} main-sequence relation at z$\sim$5. We do not observe a clear correlation in this plot.

% strong feedback and outflow mechanism that can regulate metallicity

A consequence of strong feedback is gas outflow, which can be traced with \cii emission \citep[e.g.][]{Cicone2015Very6}. Large-scale outflows \citep[e.g.][]{Bischetti2019WidespreadUniverse, Ginolfi2020TheUniverse, Pizzati2023COutflows, Romano2023Star-formation-driveniHerschel/i} may be escaping the galaxy with enriched gas that glows in \cii emission while the UV emission only traces the stars within the galaxy. Thus, the bulk of the \cii-emitting metal-enriched gas may be located in a different region than the bulk of the UV-emitting stars. This may produce an observable offset much larger than the size \citep[e.g.][]{Baron2018DirectWinds} of the star-forming regions in the galaxy. In Fig.~\ref{fig:veloff_cii_uv}, we plot the velocity offset between the \cii and ISM absorption lines as a function of \cii-UV spatial offset to check for correlation with outflow signatures. We again find no apparent trend.

\subsubsection{Morphology and kinematics}
\label{sec:morpho}
As mentioned in Sec.~\ref{sec:sample}, although we removed \cii-identified mergers based on \citet{Romano2021The5}, there may still be galaxies with complex optical, UV, and FIR continuum morphology, which would result in spatial offsets as described below.

\paragraph{Uneven star-formation}
Young and old stars are distributed differently in a galaxy \citep[e.g.][]{ElYoussoufi2019TheClouds}; young stars are located only where current star-formation is taking place, while older stars (whose population grows as young stars age) are more spread out \citep{Katz2017InterpretingReionization}. UV light would trace the brightest star-forming regions \citep[e.g.][]{Papovich2005TheGalaxies}, whereas optical emission would include a more evenly distributed older stellar population, thereby producing an offset between the two emissions.

Some galaxies may also have non-uniform or clumpy star-formation \citep[e.g.][]{Guo2012Multi-wavelength2, Hatsukade2015High-resolution3.042} in either UV or optical emission as can be seen for DEIMOS\_COSMOS\_403030 and DEIMOS\_COSMOS\_873756 in Appendix~\ref{apx:significant_offsets} Fig.~\ref{fig:off4} and ~\ref{fig:off5}. In these cases, the offset illustrates a complex morphology in one of the emissions (or undetected complexity in the remaining emissions), rather than a physical separation. %last one is more for cii-uv offset than opt-uv offset
We require higher resolution observations at longer wavelengths than UltraVISTA K-band (with e.g. \emph{JWST}) to test these scenarios by observing an even older stellar population.
% \end{itemize}

\paragraph{Differential dust attenuation}

The dust distribution across the galaxy may be non-uniform \citep{Graziani2020TheProperties, Sommovigo2020WarmImplications}, with some highly obscured and some relatively dust-free regions. In this case, the UV/optical emission from the stars within the obscured regions is almost entirely absorbed by the dust, making them invisible to \emph{HST} and VISTA. However, ALMA can still see the re-radiated light in FIR, and hence FIR continuum remains unaffected by the obscuration. Thus, we may observe an offset between the UV/optical emission probing only the dust-free regions compared to the FIR continuum emission probing obscured regions \citep[e.g.][]{Cochrane2021ResolvingScales, Hodge2016Kiloparsec-scaleGalaxies, Behrens2018DustySimulations, Rivera2018Galaxies}. In Fig.~\ref{fig:beta_cont_uv}, we plot median UV continuum slope ($\beta$; %I measured the UV slope using the Calzetti bandpass directly on Monte-Carlo'd photometry, so there was no assumption of a dust law, as there would be if I, e.g., measured on the best-fit template(s). so you could say we used the Calzetti spectral range to measure it, but we didn't assume the Calzetti attenuation law.  - BL
measured using \citet{Calzetti2000TheGalaxies} spectral range; see \cite{Faisst2020TheMeasurements}) as a function of FIR continuum-UV offset to see if offsets correlate with dust obscuration. Although, it is important to note that since the $\beta$ slope is derived from UV emission, it cannot accurately measure the dust content of highly dust-obscured galaxies. Moreover, we currently only have one galaxy on this plot with a significant offset, so we require more dust continuum observations to produce large number statistics.
%DEIMOS\_COSMOS\_881725 shows multiple components in its FIR continuum image \citep[see Fig.~D.1 of][]{Bethermin2020TheProperties}.

\paragraph{UV-dark or FIR-dark galaxies}
% Some ALPINE mergers are separated by small distances (e.g., DC818760, DC873321).
% -An interesting example here is Jekyll & Hyde, a pair of interacting galaxies at z~3. One is bright in FIR, the other in UV. https://arxiv.org/pdf/1709.03505.pdf
Considering the situation of two galaxies in a close merger (not identified as such in kinematic analyses due to the limited spatial resolution), it may be that one of them has very low dust and metallicity and hence, only emits in UV-optical \citep[e.g.][]{Ouchi2013ANOBSERVATIONS, Matthee2019ResolvedReionization, Romano2022TheRelation}, while the other is highly dust-obscured and thus, only emits in FIR \citep[e.g.][]{Bowler2018ObscuredGalaxies, Wang2019AUniverse, Romano2020TheGalaxy, Fudamoto2021NormalReionization, Talia2021IlluminatingBang, Fujimoto2022ADawn, Algera2023Cold7}. Hence, \emph{HST} will only detect the UV emitting galaxy, whereas ALMA will only detect the FIR emitting galaxy. In such a scenario, not only would we fail to identify that there are two distinct galaxies in a close merger, introducing a bias in the estimation of the real merger fraction \citep{Romano2021The5}, we would also derive a ``spatial offset'' between the UV and FIR emission from two separate galaxies. Indeed, Posses et al. (in prep.) have found evidence of such a merger in DEIMOS\_COSMOS\_683613 (Fig.~\ref{fig:sig_offset}). Higher resolution observations of this galaxy from JWST and the ALMA-CRISTAL project (Herrera-Camus et al. (in prep.)) have revealed multiple components in both UV/optical and FIR emission.

Even in galaxies without such major mergers where the derived offset is consistent with the size of a typical galaxy at $z\sim5$, i.e., a few kpc \citep[e.g.][]{Fujimoto2020Structure, Ribeiro2016SizeSurvey}, there may still be distinct regions within the same galaxy. This may be the case for a few of our $s > 3$ galaxies (see Appendix~\ref{apx:off_phys} Fig.~\ref{fig:physical_properties6}k), and may be the complex dust geometry scenario discussed above. In fact, sources where the offset is significantly larger than the galaxy size may be a different population than those where the offset is smaller than the galaxy size; the larger offsets may indicate environmental factors (e.g.: companion galaxies interacting) while smaller offsets may indicate processes internal to the galaxy (e.g.: disk instabilities that form clumps). \citet{Carniani2017ExtendedALMA} have proposed some other scenarios such as pristine gas inflows or past outflows with low metallicity and dust (hence invisible to ALMA), but with in-situ star formation (visible in UV), or accreting satellite clumps with obscured star-formation (visible in FIR) nearby a less obscured galaxy (visible in UV).

% However the merging scenario becomes more likely when the offset is larger. Since nearly all our offset estimates (see Fig.~\ref{fig:histograms}) are $\lesssim$1'', it is unlikely that our UV/optical observations and FIR observations are from different galaxies.

\paragraph{Kinematics}
\citet{Jones2021The4.4-5.9} provide a kinematic classification for a high-mass (M$_\star$ > $10^{9.5}$\,M$_\odot$) subset of the ALPINE sample, using various methods: the tilted ring model fitting code $^\mathrm{3D}$Barolo \citep{Teodoro20153DGalaxies}, morphological classification with Gini-M20 \citep{Lotz2004AClassification}, and several disk identification criteria \citep{Wisnioski2015The2.7}. Of the 17 galaxies in our significant offset sample, seven have a classification from \citet{Jones2021The4.4-5.9}, with three rotation-dominated (ROT), two dispersion-dominated (DIS), and two uncertain (UNC). The galaxies with ROT classification show significant but small offsets in our analysis (such as DEIMOS\_COSMOS\_396844 in Appendix~\ref{apx:significant_offsets} Fig.~\ref{fig:off1}). This may indicate a hidden complex morphology that is not discernible at the current resolution. DEIMOS\_COSMOS\_873756 (Appendix~\ref{apx:significant_offsets} Fig.~\ref{fig:off5}), classified as DIS, has multiple significant offsets, between optical and all other emissions. DEIMOS\_COSMOS\_683613 (Fig.~\ref{fig:sig_offset}), classified as UNC, also has multiple offsets between FIR continuum and all other emissions. However, as discussed in the previous section, this galaxy has been identified as a merging system with higher resolution (Posses et al. (in prep.)). Moreover, of the 37 galaxies in our sample with no significant offsets, three have been classified as ROT, one as DIS and eight as UNC. DEIMOS\_COSMOS\_881725, despite having no significant offset and ROT classification, shows multiple components in its FIR continuum image \citep[see Fig.~D.1 of][]{Bethermin2020TheProperties}. It is therefore difficult to find any clear relationship between kinematic classification and significant offsets.

\subsubsection{Galaxy orientation}
% check references: https://iopscience.iop.org/article/10.3847/1538-4357/abec76/meta
% https://academic.oup.com/mnras/article/458/3/2443/2589295

The orientation of the galaxy on the sky may amplify the effects of uneven dust distribution \citep[e.g.][]{Yip2010EXTINCTIONSPECTRA, Devour2017RevealingStructures}. Some galaxies may be oriented on the sky such that we can directly observe the inner star-forming regions, e.g.: a face-on spiral with dust distributed evenly across the disk. Other galaxies might be partially dust-obscured from our point of view, e.g: an edge-on spiral with a dusty disk obscuring part of the central bulge. Whereas in the former case, centroids of the UV/optical emission and FIR continuum emission will be co-spatial, in the latter case, the UV/optical emission will arise from the unobscured part of the disk, while the FIR continuum centroid may be located near the part of the disk with the highest concentration of dust. Then, although both sources are physically the same kind of galaxy, we would interpret them as different kinds of objects based on the offset. We may be able to quantify the effects of galaxy orientation with higher spectral resolution by studying the \cii line profile \citep[e.g.][]{Kohandel2019KinematicsEmission}. %, or by observing ALPINE galaxies across many wavelengths from UV to FIR.
That said, in our case, offset contribution from orientation effects is likely small given our resolution.

Overall, none of the plots in Fig.~\ref{fig:scatter} show clear trends. Several more plots are presented in Appendix~\ref{apx:off_phys}, but in all cases, either the number of galaxies with significant offsets is too small to observe a correlation, or there is no definitive trend. %\tcr{we would naively expect that offsets would correlate with sfr, dev MS, vel off, etc. but they don't => this is a result in itself}
Therefore, understanding which of these scenarios is driving the spatial offsets requires observations \citep[e.g.][; Posses et al. (in prep.)]{Herrera-Camus2021AProperties, Chen2022JWST/NIRCamScales} or simulations \citep[e.g.][]{Graziani2020TheProperties, Rizzo20227, Pallottini2022ASimulations} that resolve galaxies down to sub-arcsecond scales.
%\cite cristal if available before publishing this paper

\subsection{Consequences of significant spatial offsets}

\begin{figure}
    \centering
    \includegraphics[width=\columnwidth]{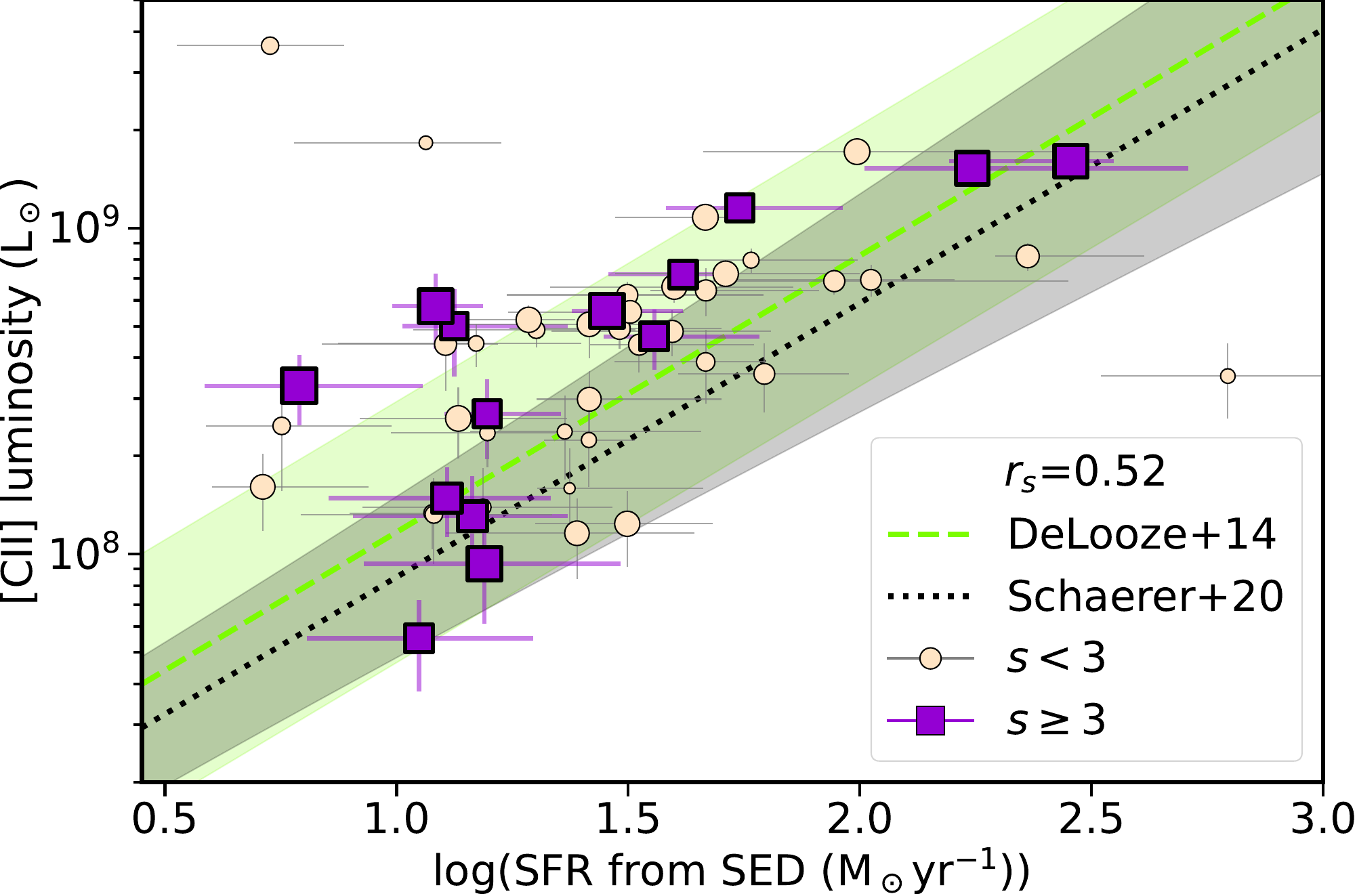}
    \caption{The $L_{\rm [CII]}$-SFR relation for galaxies with and without significant spatial offsets. As in Fig.~\ref{fig:scatter}, the galaxies with a significant \cii-UV offset are shown as violet squares with errorbars in the same colour, while those without significant offsets are shown as cream-coloured circles with errorbars in grey. Markersize increases with significance. The relations from \citet{DeLooze2014TheTypes} and \citet{Schaerer2020TheGyr} are plotted as green dashed and black dotted lines respectively. The corresponding uncertainty in these relations is shown as shaded regions in the same colour.}
    \label{fig:cii_sfr}
\end{figure}

The prevalence of significant spatial offsets in high-$z$ galaxies may affect many commonly used relations at these redshifts. For instance, significant offsets between \cii and UV emission could alter the \cii-SFR relation \citep{Schaerer2020TheGyr, Romano2022TheRelation, Ferrara2022The7} as the \cii emission would arise from gas that is tracing a different region, away from the site of star formation. In Fig.~\ref{fig:cii_sfr}, we plot the $L_{\rm [CII]}$-SFR relation and highlight the galaxies with significant \cii-UV offsets. We see that the galaxies with significant offsets tend to lie above those without, thereby affecting the overall relation. This may indicate that the SED-derived SFR is underestimated for galaxies with offsets, possibly due to missing or undetected emission.

Similarly, FIR-UV offsets would affect the IRX-$\beta$ relation \citep[e.g.][]{Faisst2017Properties,Popping2017DissectingGalaxies, Narayanan2018TheSimulations,Fudamoto2020TheZ4.4-5.8, Sommovigo2020WarmImplications, Boquien2022The4.4-5.5}. UV emission is used to get IRX-$\beta$, which is then assumed to be the dust content in FIR, but if emission does not originate in the same physical region in the galaxy \citep[e.g.][]{Casey2014AreGalaxies, Gomez-Guijarro20184.5, Elbaz2018StarburstsSequence}, this assumption does not hold.

Many galaxy SED modelling codes (e.g.\ CIGALE \citep{Burgarella2005StarAnalysis, Noll2009AnalysisSample, Boquien2019CIGALE:Emission, Pacifici2023TheTechniques}, MAGPHYS \citep{DaCunha2008AGalaxies}) assume an energy balance between UV and FIR emission, which may not hold in the presence of offsets, thereby affecting SED-derived estimates of stellar mass and SFR. \citet{Ferrara2022The7} define a dimensionless quantity called the molecular index, which is a ratio between the IR-to-UV continuum flux and the difference between observed  and intrinsic $\beta$ slopes. This quantity measures the extent to which IR and UV emission are decoupled. \citet{Sommovigo2022AGalaxies} study the ALPINE sample and find that SED derived SFRs do not match the total (including UV and IR) SFRs for galaxies that have a high molecular index, which they interpret as a consequence of spatially decoupled UV and IR emission. The galaxy with the largest discrepancy in their analysis (DEIMOS\_COSMOS\_873756; shown in Appendix~\ref{apx:significant_offsets} Fig.~\ref{fig:off5}) is one of the galaxies we find to have a significant spatial offset. Using higher resolution observations, \citet{Mitsuhashi2023TheZ=4-6} similarly find a significant difference between SED-derived SFR and total SFR of this galaxy (called CRISTAL-24 in their sample). Our results lend support to the idea put forth by these works of spatially decoupled IR and UV emission in this galaxy.

On the other hand, \citet{Haskell2023EnergyOffsets} tested the effect of offsets with MAGPHYS, and found that offsets, however large, have no appreciable impact on SED-derived properties (provided an acceptable fit is obtained) in over 80 percent of their sample. They propose that the underlying \citet{Charlot2000AGalaxies} dust model has sufficient flexibility to account for the differential dust attenuation between the decoupled UV-bright and FIR-bright regions of the galaxies with offsets.

In any case, deviations (if any) caused by spatial offsets can be mitigated via spatially resolved SED modelling \citep[e.g.][]{Wuyts2012SmoothERGalaxies, Sorba2018SpatiallyProblem}.% Hence, the non co-spatial nature of emission should be accounted for in SED modelling to derive accurate physical properties.

The presence of offsets may also affect follow-up ALMA observations of JWST targets (or vice-versa). If there is a spatial offset, this should be considered when planning observations and interpreting data. For instance, JWST/NIRSpec has a slit width of only $\sim0.2$\,arcsec \citep{Jakobsen2022TheTelescope}, comparable to our median uncertainties, and much smaller than our median significant offsets (see Table~\ref{tab:subsamples_offs}). Large surveys with this instrument, e.g.\ Cosmic Evolution Early Release Science \citep[CEERS][]{Finkelstein2023CEERSJWST}, JWST Advanced Deep Extragalactic Survey \citep[JADES][]{Eisenstein2023OverviewJADES}, might only observe the UV emission, and miss the dusty component. Therefore, spatial offsets must be taken into account for all studies, not just spatially resolved galaxy modelling.

% \tcr{more consequences?}

\section{Summary and Conclusions}
We study a sample of main-sequence star-forming galaxies at \(z \sim 4\text{--}6\) from the ALPINE dataset. We identify 54 galaxies that were detected in \cii and FIR continuum emission in ALMA data, UV emission in \emph{HST} data, and optical emission in K-band UltraVISTA data, excluding \cii-mergers or multi-component systems based on \citet{Romano2021The5} and \citet{Jones2021The4.4-5.9}.

We use the \emph{HST} coordinates (with astrometric correction) from \citet{Faisst2020TheMeasurements} as the UV centroids. To calculate \cii, FIR continuum, and optical centroids, we fit 2D Gaussians to the detected emission and apply a bootstrapping technique to estimate fit centroids and uncertainties. We convolve the positional accuracy of the respective telescopes and ALMA noise correlation uncertainty with the fit uncertainty to calculate the total uncertainty. We then estimate the spatial offset between centroids of detected emission for each galaxy, along with offset significance ($s$), which is calculated as the measured offset divided by the total uncertainty on the offset.

\begin{itemize}
    \item The (sigma-clipped) median of the measured offsets is 0.1--0.3\,arcsec, which translates to \(\sim\)0.6--2\,kpc at our median redshift of \(z\sim5\).
    \item We establish a cut-off of $s>3$ to define ``significant'' offsets. These significant offsets are \(\sim\)0.5--0.7\,arcsec, or \(\sim\)3.5--4.5 kpc.
    \item 17 galaxies (\(\sim\)30 percent of the sample) display significant offsets between one or more emission pairs, although none have all four emissions offset from each other. The remaining 37 galaxies (\(\sim\)70 percent of the sample) have no significant offsets.
    \item  We discuss several potential phenomena that may lead to the observation of spatial offsets, plotting corresponding galaxy physical properties against their measured spatial offsets wherever possible. We find no clear trends or the statistics are too low to make strong conclusions. The physical origin of the observed offsets is therefore still unclear.
\end{itemize}

The existence of significant spatial offsets in \(\sim\)30 percent of our sample indicates that it is possible for main-sequence galaxies at \(z\sim4\text{--}6\) to have the bulk of the stars spatially offset from the bulk of the interstellar medium. Future simulations and observations must therefore take into account that the emission observed across wavelengths may be coming from different, spatially segregated regions of the galaxy.%As this picture runs counter to the assumption of co-spatial emission in galaxy SED fitting codes,

We require large number statistics and higher resolution observations and simulations to identify the processes driving spatial offsets. For instance, we could perform this analysis on the REBELS sample \citep[already shown to have spatial offsets in][]{Inami2022The6.5}, which has different SFRs and M$_\star$, but similar angular resolution as ALPINE. \emph{JWST}, with its superior angular resolution, may also be able to shed light on the origin of offsets \citep[e.g.][]{Bakx2023DeepCandidate}.

\section*{Acknowledgements}

We thank the anonymous referee for insightful comments, and Kate Whitaker, Francesca Rizzo, and Ana Posses for useful discussions.
%referee
The Cosmic Dawn Center is funded by the Danish National Research Foundation under grant number 140. M.K. was supported by the ANID BASAL project FB210003.
%MK
This publication has received funding from the European Union’s Horizon 2020 research and innovation programme under grant agreement No 101004719 (ORP). 
%related to alpine funding
M.R. acknowledges support from the Narodowe Centrum Nauki (UMO-2020/38/E/ST9/00077) and support from the Foundation for Polish Science (FNP) under the program START 063.2023. E.I. acknowledges funding by ANID FONDECYT Regular 1221846. GCJ acknowledges funding from the ``FirstGalaxies'' Advanced Grant from the European Research Council (ERC) under the European Union’s Horizon 2020 research and innovation programme (Grant agreement No. 789056). M.B. acknowledges support from FONDECYT regular grant 1211000 and by the ANID BASAL project FB210003. 021-001131-S funded by
MCIN/AEI/ 10.13039/501100011033. H.M.H. acknowledges support from National Fund for Scientific and Technological Research of Chile (FONDECYT) through grant no. 3230176

This paper is based on data obtained with the ALMA Observatory, under Large Program 2017.1.00428.L. ALMA is a partnership of ESO (representing its member states), NSF (USA) and NINS (Japan), together with NRC (Canada), MOST and ASIAA (Taiwan), and KASI (Republic of Korea), in cooperation with the Republic of Chile. The Joint ALMA Observatory is operated by ESO, AUI/NRAO and NAOJ. 
%alpine
This paper makes use of archival data from the NASA/ESA \emph{Hubble Space Telescope}. Based on data obtained with the European Southern Observatory Very Large Telescope, Paranal, Chile, under Large Program 185.A-0791, and made available by the VUDS team at the CESAM data centre, Laboratoire d’Astrophysique de Marseille, France. This work is based on observations taken by the 3D-HST Treasury Program (GO12177 and 12328) with the NASA/ESA \emph{HST}, which is operated by the Association of Universities for Research in Astronomy, Inc., under NASA contract NAS5-26555. 
%hst - correct?
Based on data products from observations made with ESO Telescopes at the
La Silla Paranal Observatory under ESO programme ID 179.A-2005 and on data products produced by TERAPIX and the Cambridge Astronomy Survey Unit on behalf of the UltraVISTA consortium.
%ultravista - correct?
% \tcm{, and Shengqi Yang and Luca Di Mascolo for helpful discussions.}

% This paper makes use of the following ALMA data: ADS/JAO.ALMA\#2013.1.01064.S, ADS/JAO.ALMA\#2015.1.01406.S, ADS/JAO.ALMA\#2016.1.00954.S, ADS/JAO.ALMA\#2017.1.00775.S, ADS/JAO.ALMA\#2019.1.01778.S. ALMA is a partnership of ESO (representing its member states), NSF (USA) and NINS (Japan), together with NRC (Canada), MOST and ASIAA (Taiwan), and KASI (Republic of Korea), in cooperation with the Republic of Chile. The Joint ALMA Observatory is operated by ESO, AUI/NRAO and NAOJ. The National Radio Astronomy Observatory is a facility of the National Science Foundation operated under cooperative agreement by Associated Universities, Inc.

% DW and SF are supported in part by Independent
% Research Fund Denmark grant DFF-7014-00017. FR acknowledges support from the European Union’s Horizon 2020 research and innovation program under the Marie Sklodowska-Curie grant agreement No. 847523 ‘INTERACTIONS’ 

\section*{Data Availability}
The data used in the paper are available in the ALMA archive at https://almascience.nrao.edu under the program ID 2017.1.00428.L. The derived data and models generated in this research will be shared on reasonable request to the corresponding author.



\bibliographystyle{mnras}
\bibliography{00} % if your bibtex file is called example.bib

%\begin{thebibliography}{99}
%\end{thebibliography}

%%%%%%%%%%%%%%%%%%%%%%%%%%%%%%%%%%%%%%%%%%%%%%%%%%

%%%%%%%%%%%%%%%%% APPENDICES %%%%%%%%%%%%%%%%%%%%%

\appendix

\section{Galaxies with significant offsets}
\label{apx:significant_offsets}
\begin{figure*}
     \centering
     \begin{subfigure}[b]{0.28\textheight}
         \centering
         \includegraphics[width=\textwidth]{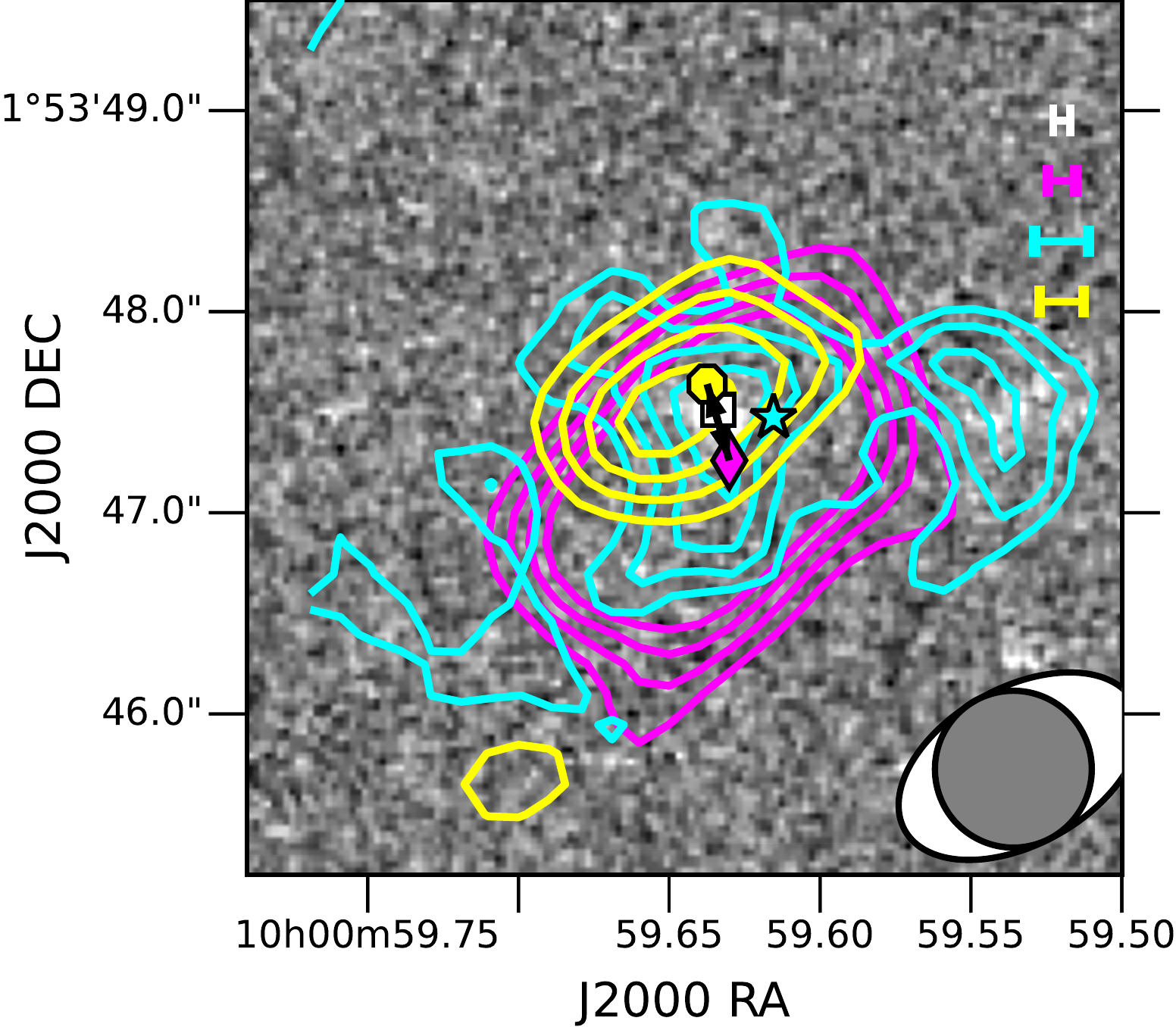}
         \caption{Significant \cii-FIR continuum offset in DEIMOS\_COSMOS\_396844}
         \label{fig:off1}
     \end{subfigure}
     \hfill
     \begin{subfigure}[b]{0.28\textheight}
         \centering
         \includegraphics[width=\textwidth]{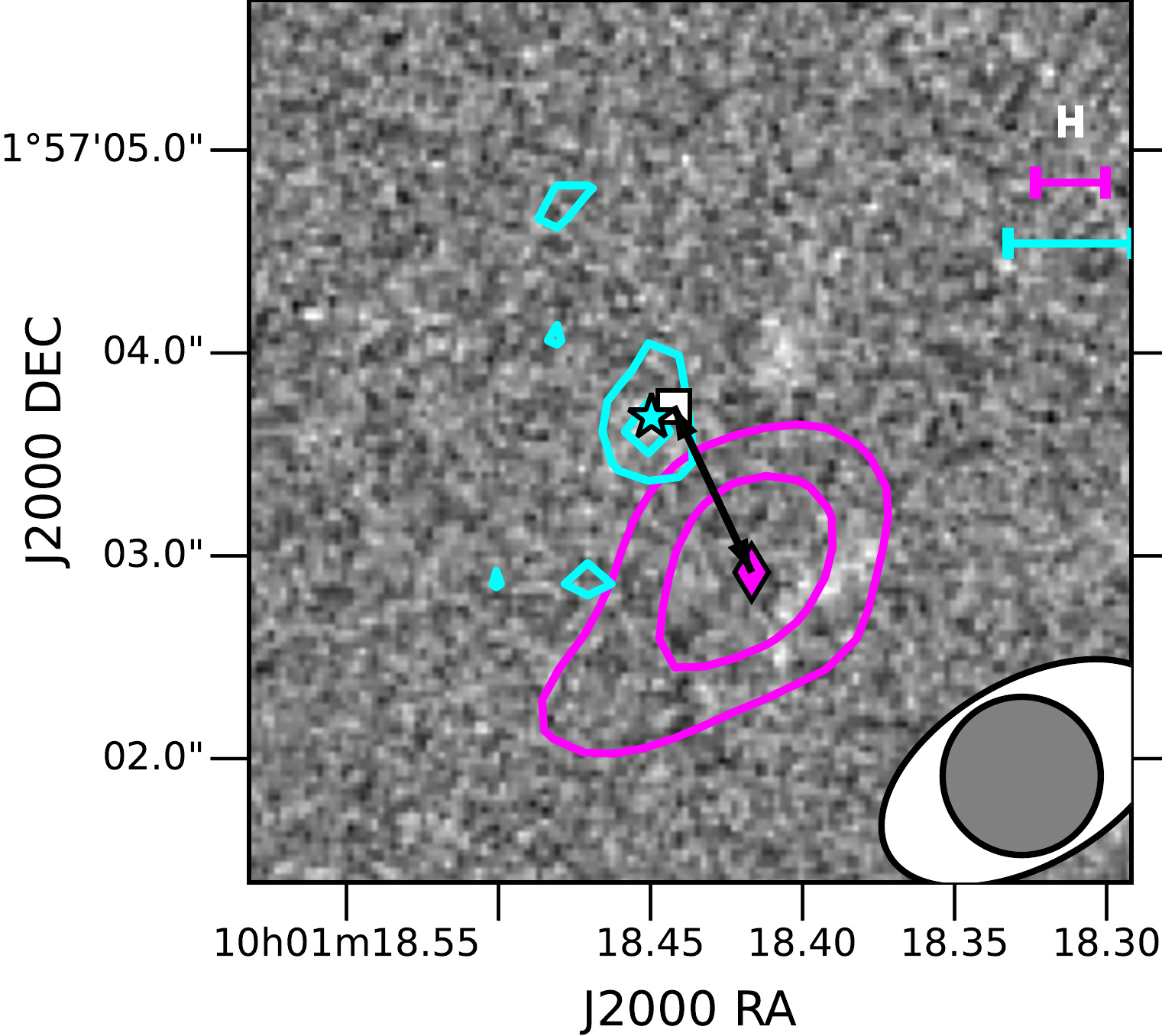}
         \caption{Significant \cii-UV offset in DEIMOS\_COSMOS\_430951}
         \label{fig:off2}
     \end{subfigure}
     \medskip
     \begin{subfigure}[b]{0.28\textheight}
         \centering
         \includegraphics[width=\textwidth]{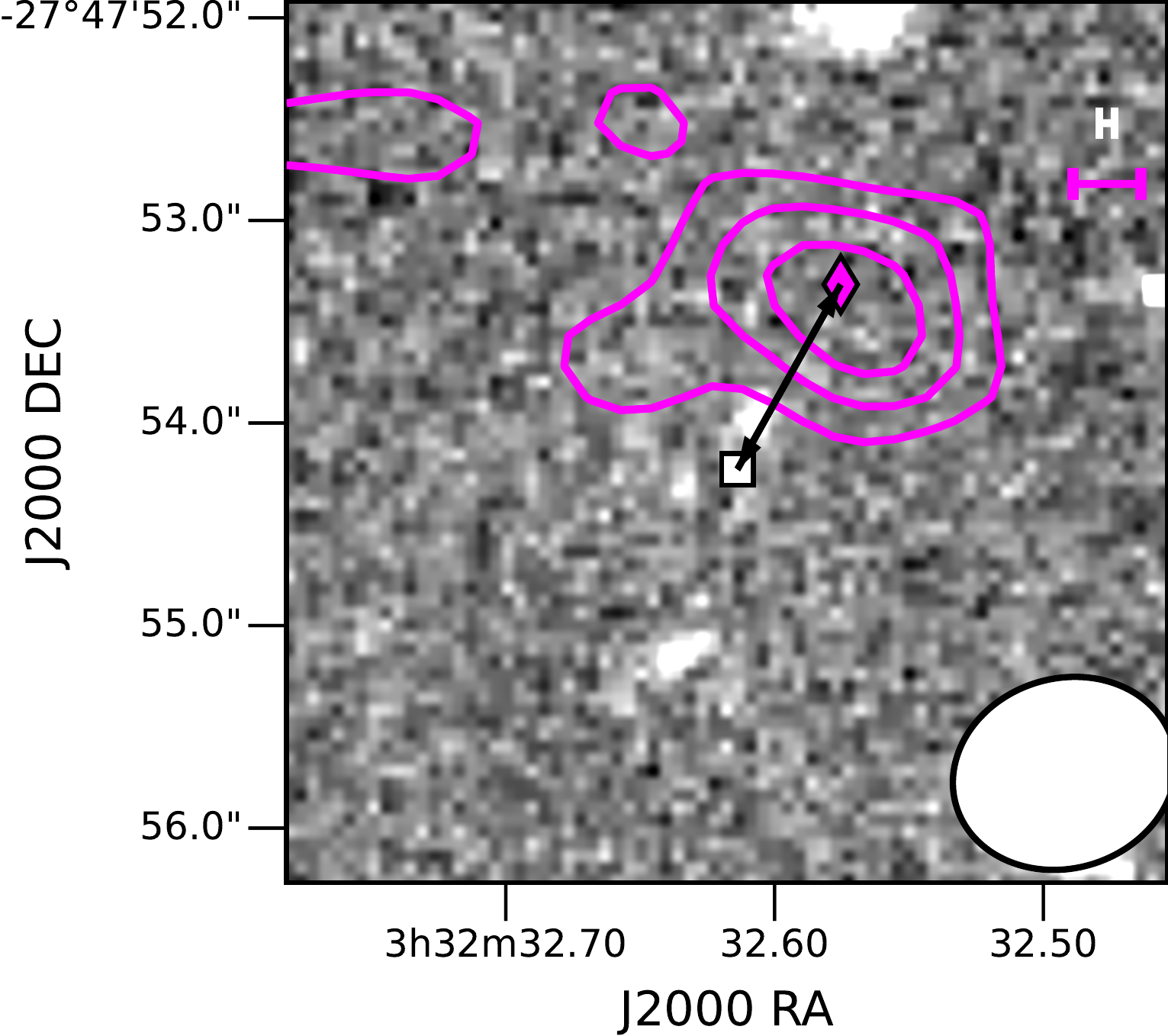}
         \caption{Significant \cii-UV offset in CANDELS\_GOODSS\_75}
         \label{fig:off3}
     \end{subfigure}
     \hfill
     \begin{subfigure}[b]{0.28\textheight}
         \centering
         \includegraphics[width=\textwidth]{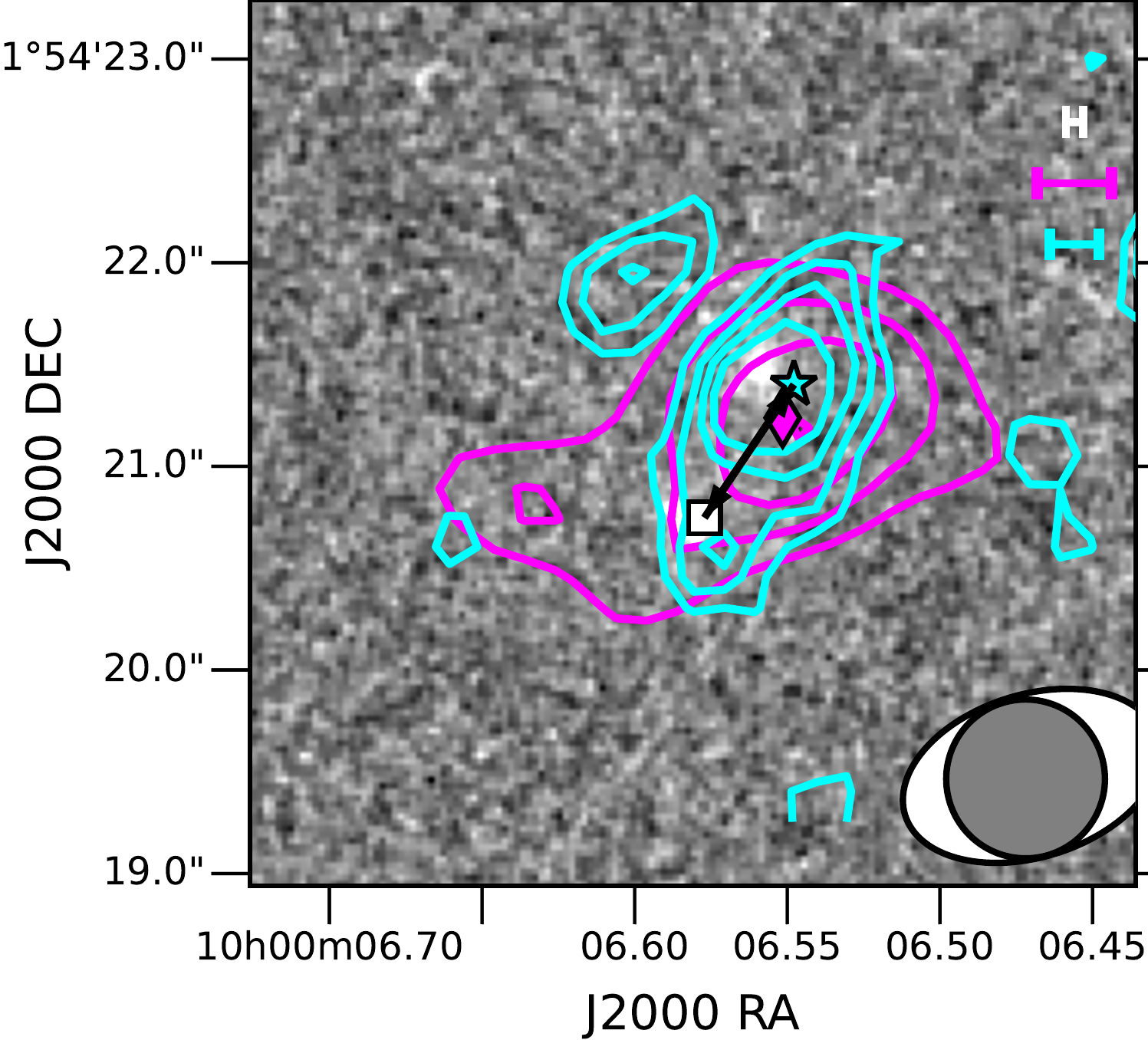}
         \caption{Significant Optical-UV offset in DEIMOS\_COSMOS\_403030}
         \label{fig:off4}
     \end{subfigure}
     \medskip
     \begin{subfigure}[b]{0.28\textheight}
         \centering
         \includegraphics[width=\textwidth]{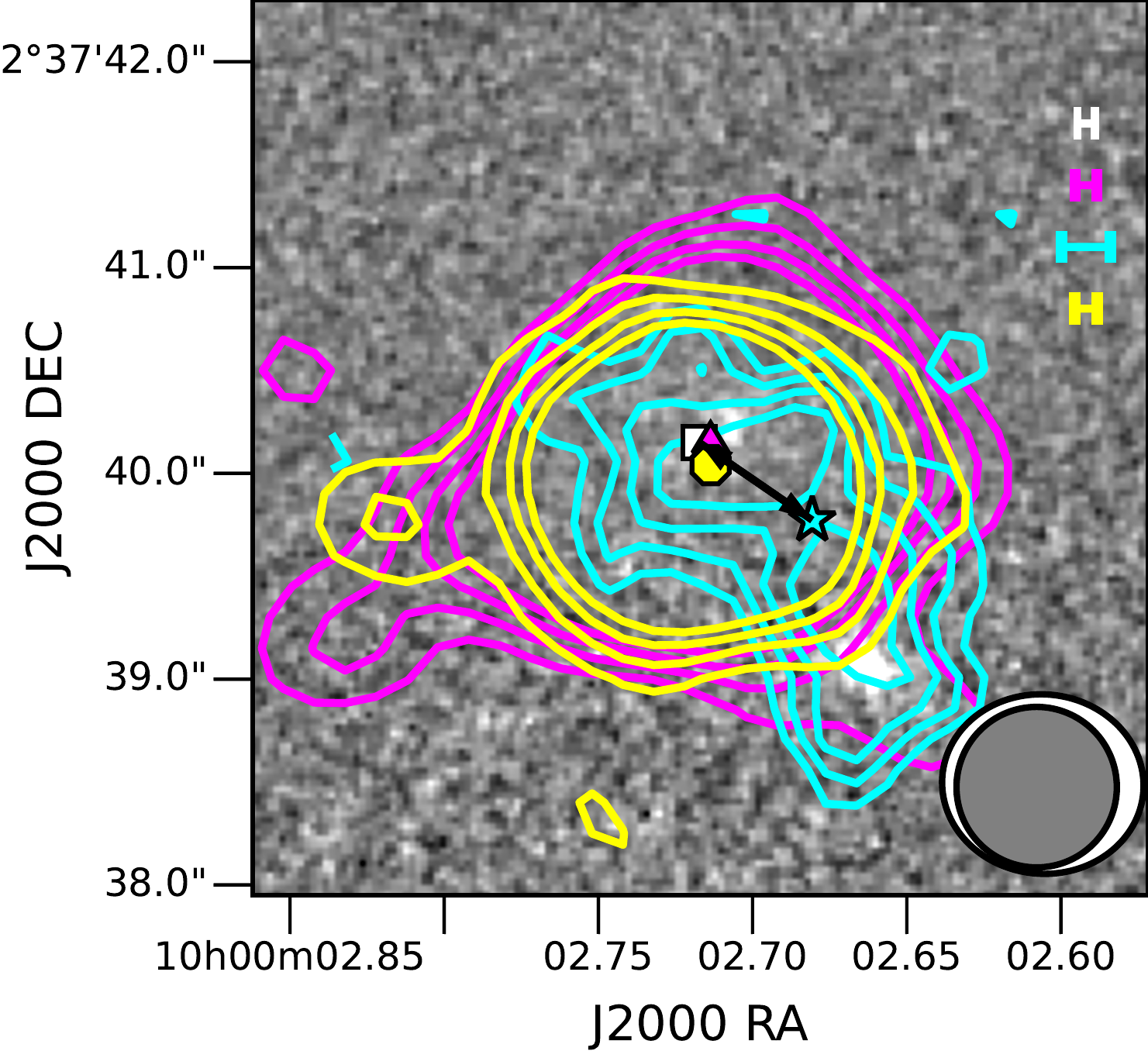}
         \caption{Significant Optical-UV offset in DEIMOS\_COSMOS\_873756}
         \label{fig:off5}
     \end{subfigure}
     \hfill
     \begin{subfigure}[b]{0.28\textheight}
         \centering
         \includegraphics[width=\textwidth]{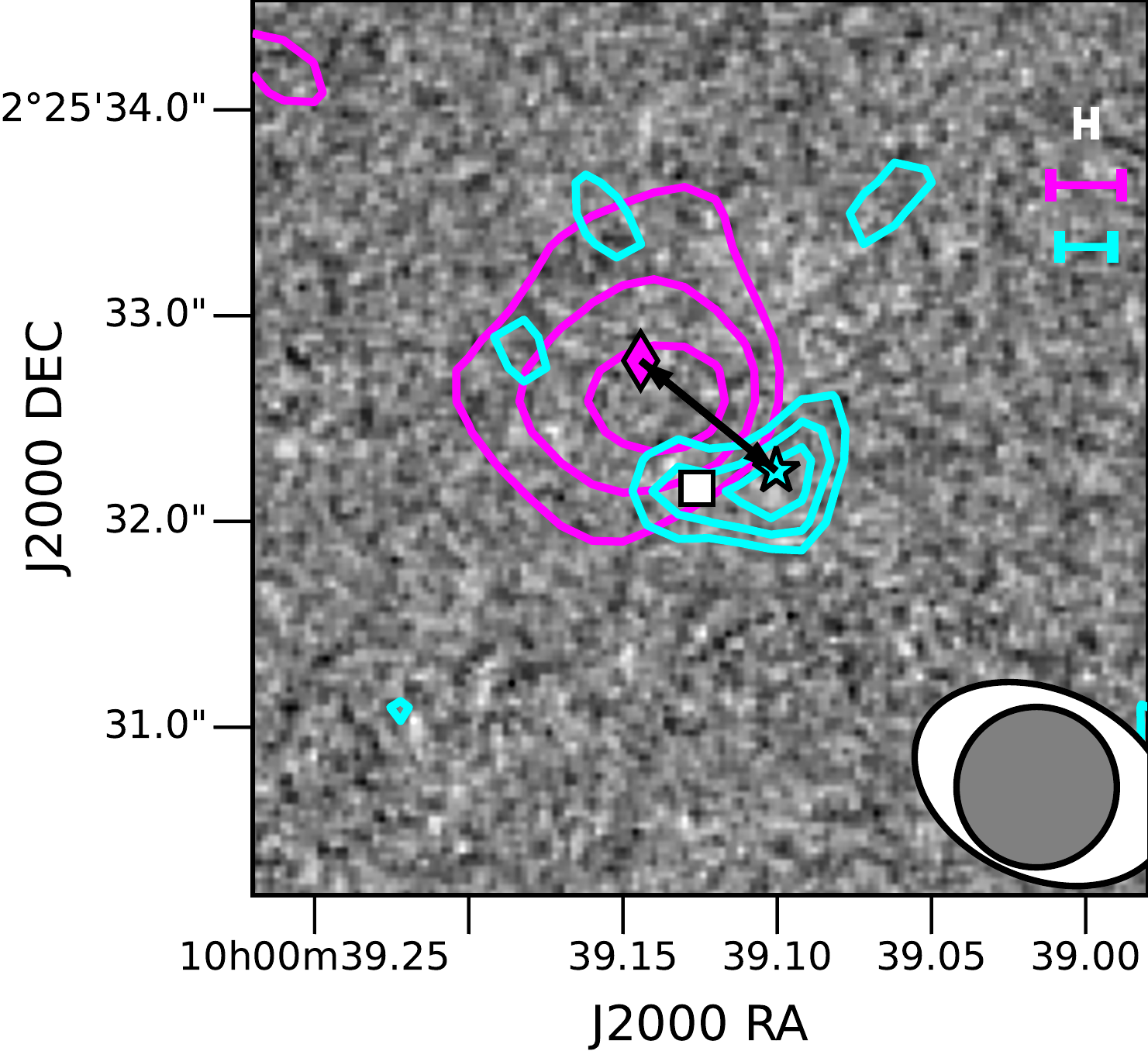}
         \caption{Significant Optical-\cii offset in DEIMOS\_COSMOS\_742174}
         \label{fig:off6}
     \end{subfigure}
\caption{Galaxies with significant offsets. Colour scheme is the same as Fig.~\ref{fig:overlaid_contours}. UV \emph{HST} image is shown as a grey-scale background with \cii (fuchsia), optical (cyan), and FIR continuum (yellow) overlaid. The contours are drawn at 2, 3, 4, and 5 times the standard deviation. The centroids are marked with a white square for UV emission, fuchsia diamond for \cii, cyan star for optical, and yellow octagon for FIR continuum (same colours as the respective contours), and the spatial offset among them is indicated by a black double-headed arrow. The calculated total error in each emission is indicated on the top right in the same colour as the contours. The ALMA (\cii and FIR continuum) beam is shown as a filled white ellipse, and optical beam as a filled grey circle.}
\label{fig:overlaid_contours2}
\end{figure*}

In Fig.~\ref{fig:overlaid_contours2}, we show several galaxies from our sample for which we measure significant offsets. UV images are shown in grey-scale, and \cii, optical, and FIR continuum (where available) are over-plotted as coloured contours. The colour scheme is the same as Fig.~\ref{fig:overlaid_contours} in the main text.
% \fig{overlaid_contours_DEIMOS_COSMOS_742174.pdf}{0.4\textwidth}{(a)}

\section{Offsets vs selected physical properties}
\label{apx:off_phys}
In Figs.~\ref{fig:physical_properties}--\ref{fig:physical_properties6}, we plot our measured spatial offsets among \cii, UV, optical, and FIR continuum emission against a range of galaxy physical properties. Colour scheme is the same as in Fig.~\ref{fig:scatter} in the main text. Galaxies with significant offsets do not display any clear trends with any of the physical properties, compared to those without significant offsets.
\begin{figure*}
     \centering
     \begin{subfigure}[b]{0.95\textwidth}
         \centering
         \includegraphics[width=\textwidth]{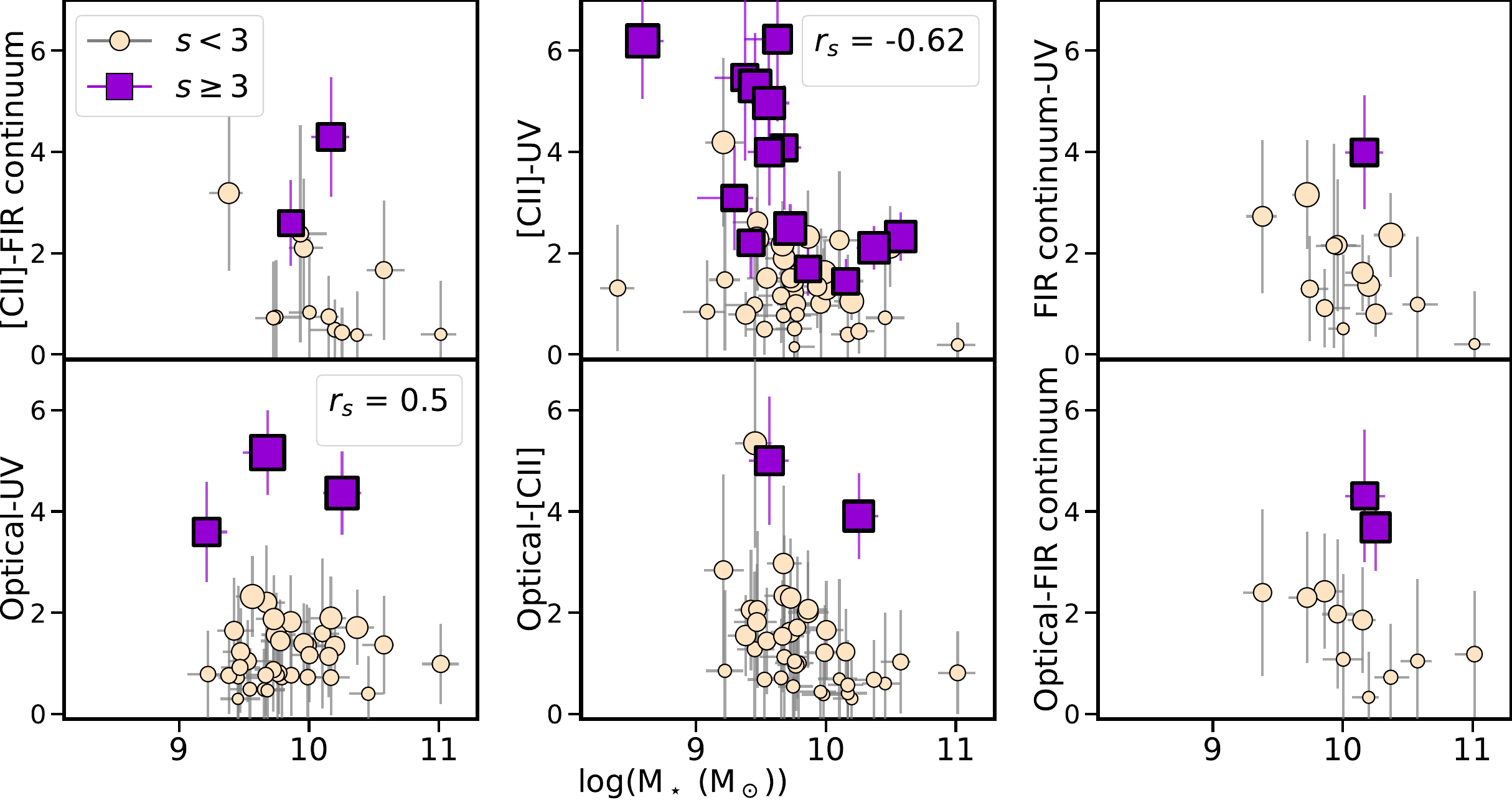}
         \label{fig:multi_mstar}
     \end{subfigure}
     \medskip
     \begin{subfigure}[b]{0.95\textwidth}
         \centering
         \includegraphics[width=\textwidth]{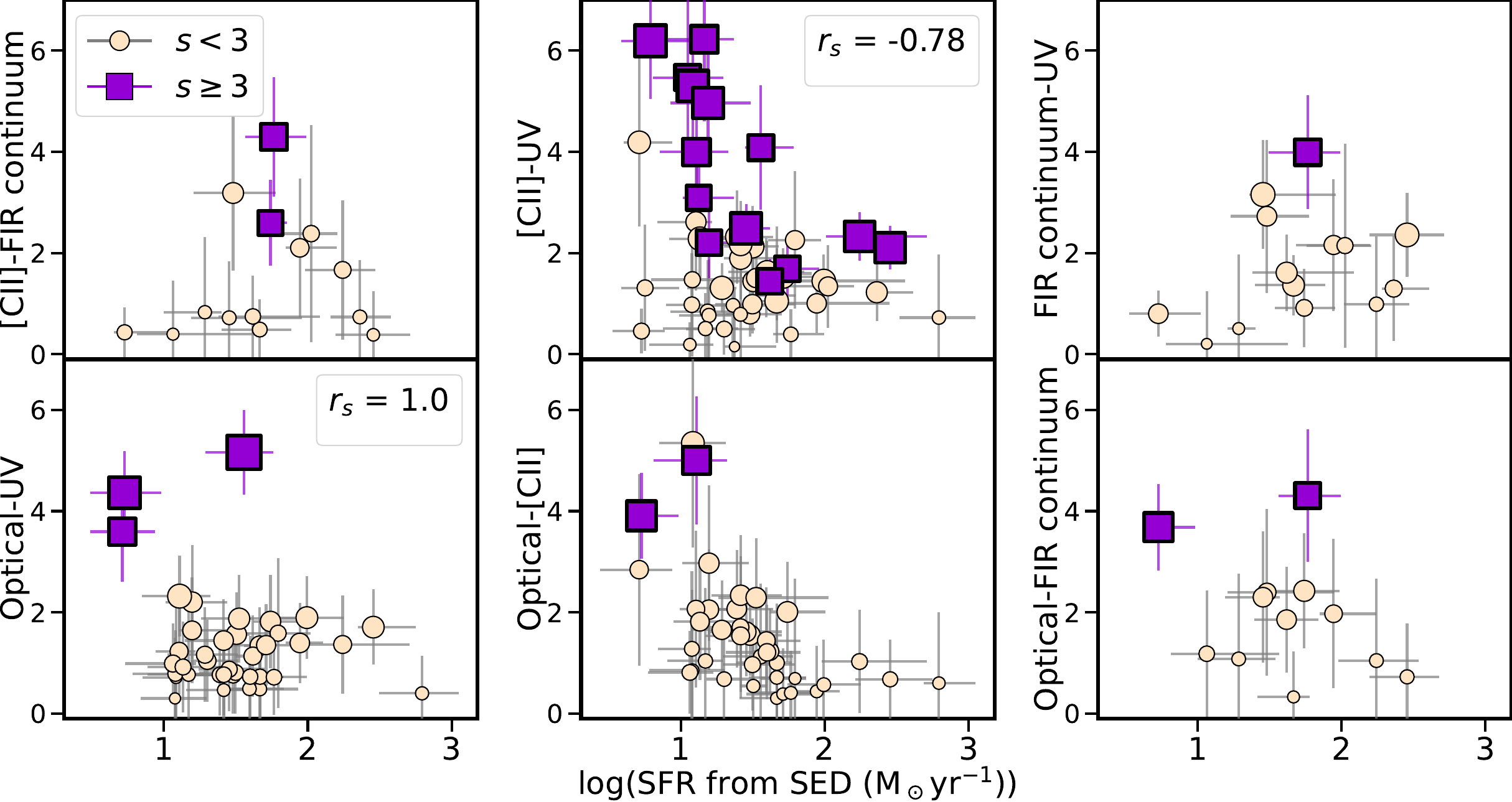}
         \label{fig:multi_sfr}
     \end{subfigure}
\caption{Offsets (in kpc) vs physical properties a) log(Stellar mass) b) Total SFR. Colour scheme is the same as in Fig.~\ref{fig:scatter}. Galaxies with significant offsets are shown as purple squares, while those without significant offsets are shown as cream-coloured circles. Marker-size increases with significance. The Spearman's rank coefficient ($r_s$) is given whenever there are three or more galaxies with significant offsets.}
\label{fig:physical_properties}
\end{figure*}

\begin{figure*}
\ContinuedFloat
\centering
     \begin{subfigure}[b]{0.95\textwidth}
         \centering
         \includegraphics[width=\textwidth]{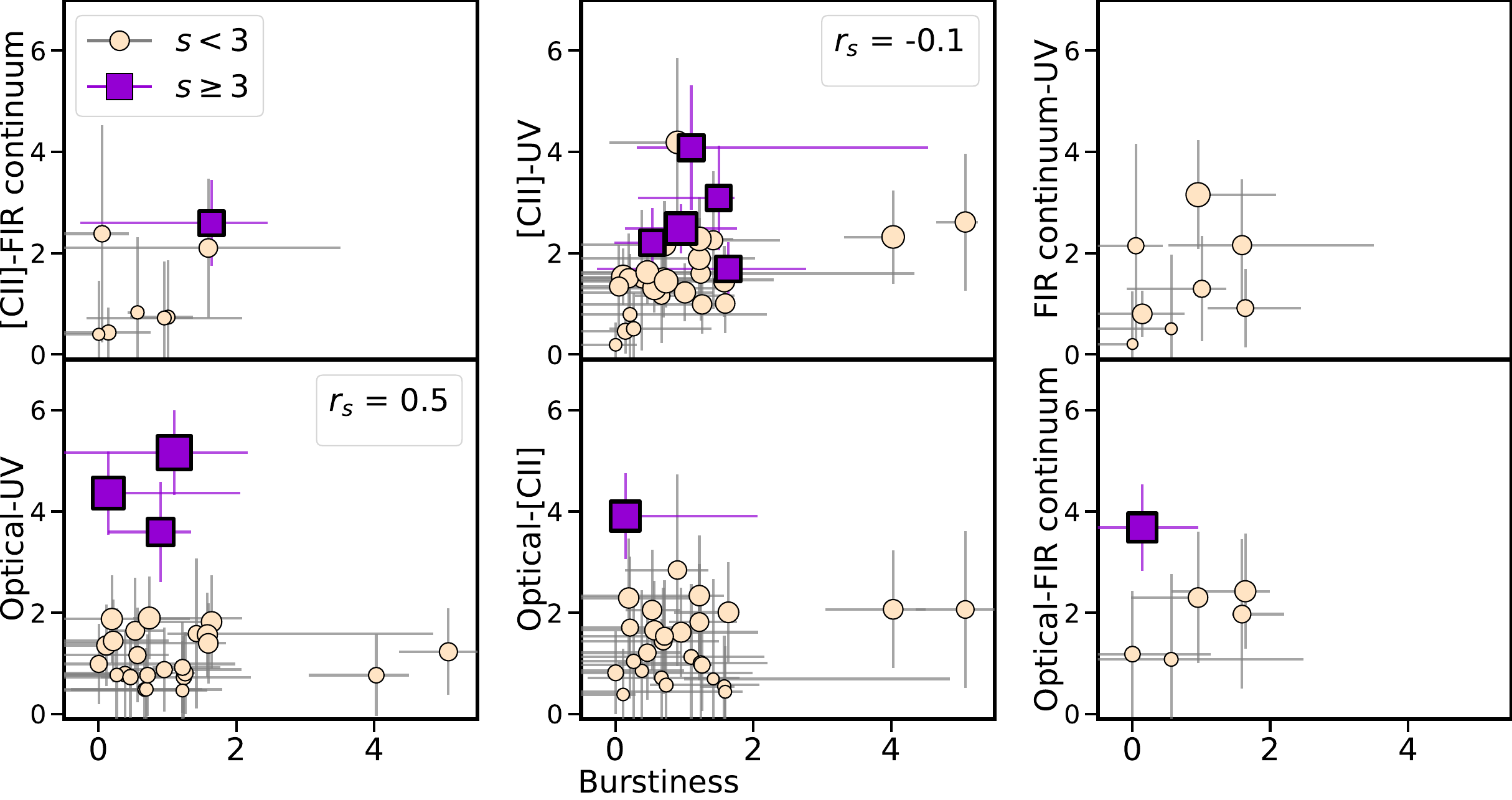}
         \label{fig:multi_burst}
     \end{subfigure}
     \medskip
     \begin{subfigure}[b]{0.95\textwidth}
         \centering
         \includegraphics[width=\textwidth]{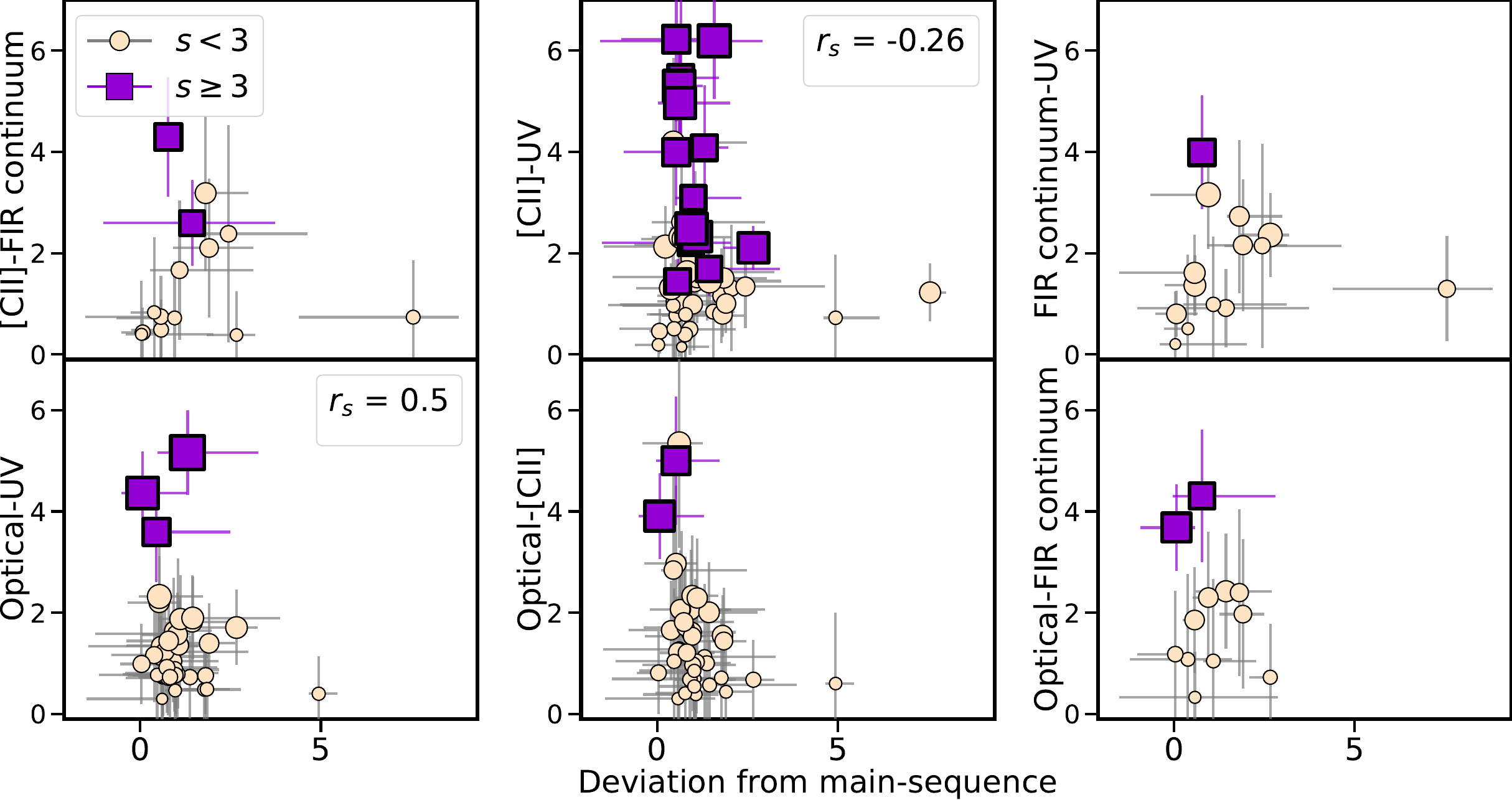}
         \label{fig:multi_devMS}
     \end{subfigure}
\caption{(contd.) c) Burstiness (as in Fig.~\ref{fig:burst_cii_uv}) d) deviation from main-sequence (as in Fig.~\ref{fig:devMS_cii_uv})}
\label{fig:physical_properties1}
\end{figure*}

\begin{figure*}
\ContinuedFloat
\centering
     \begin{subfigure}[b]{0.95\textwidth}
         \centering
         \includegraphics[width=\textwidth]{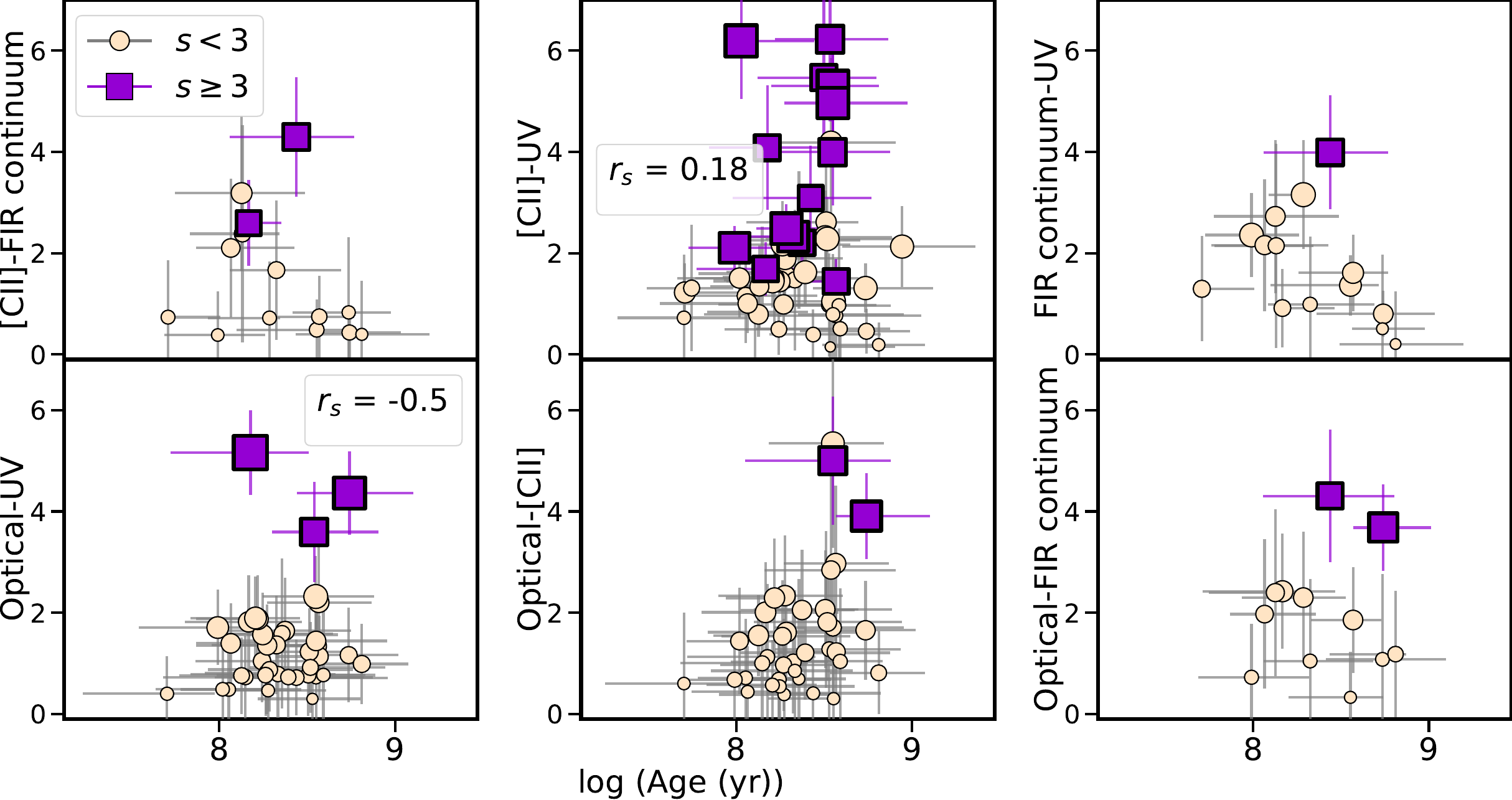}
         \label{fig:multi_age}
     \end{subfigure}
     \medskip
     \begin{subfigure}[b]{0.95\textwidth}
         \centering
         \includegraphics[width=\textwidth]{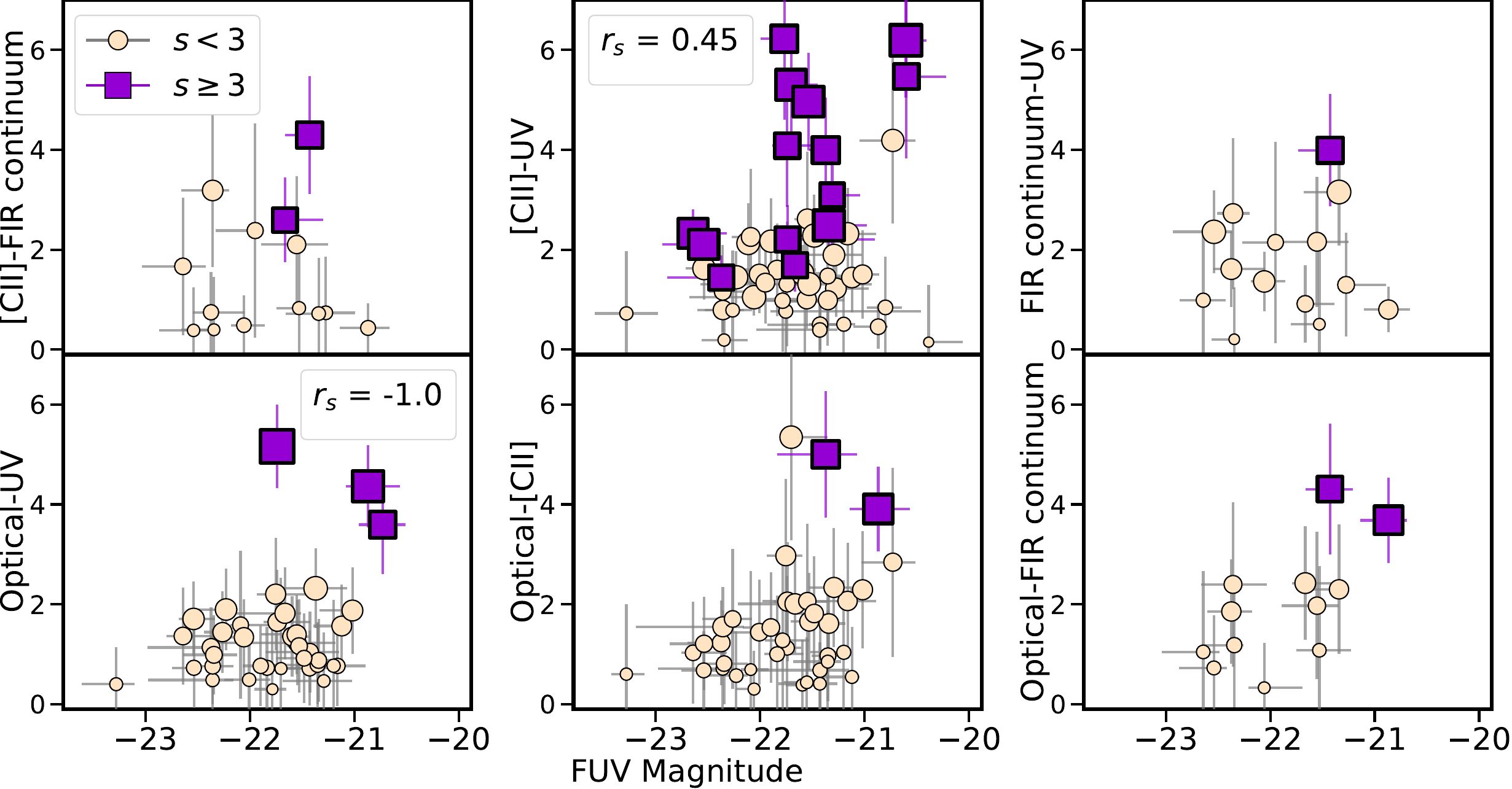}
         \label{fig:multi_mag}
     \end{subfigure}
\caption{(contd.) e) log(Age) f) FUV magnitude}
\label{fig:physical_properties2}
\end{figure*}

\begin{figure*}
\ContinuedFloat
     \begin{subfigure}[b]{0.95\textwidth}
         \centering
         \includegraphics[width=\textwidth]{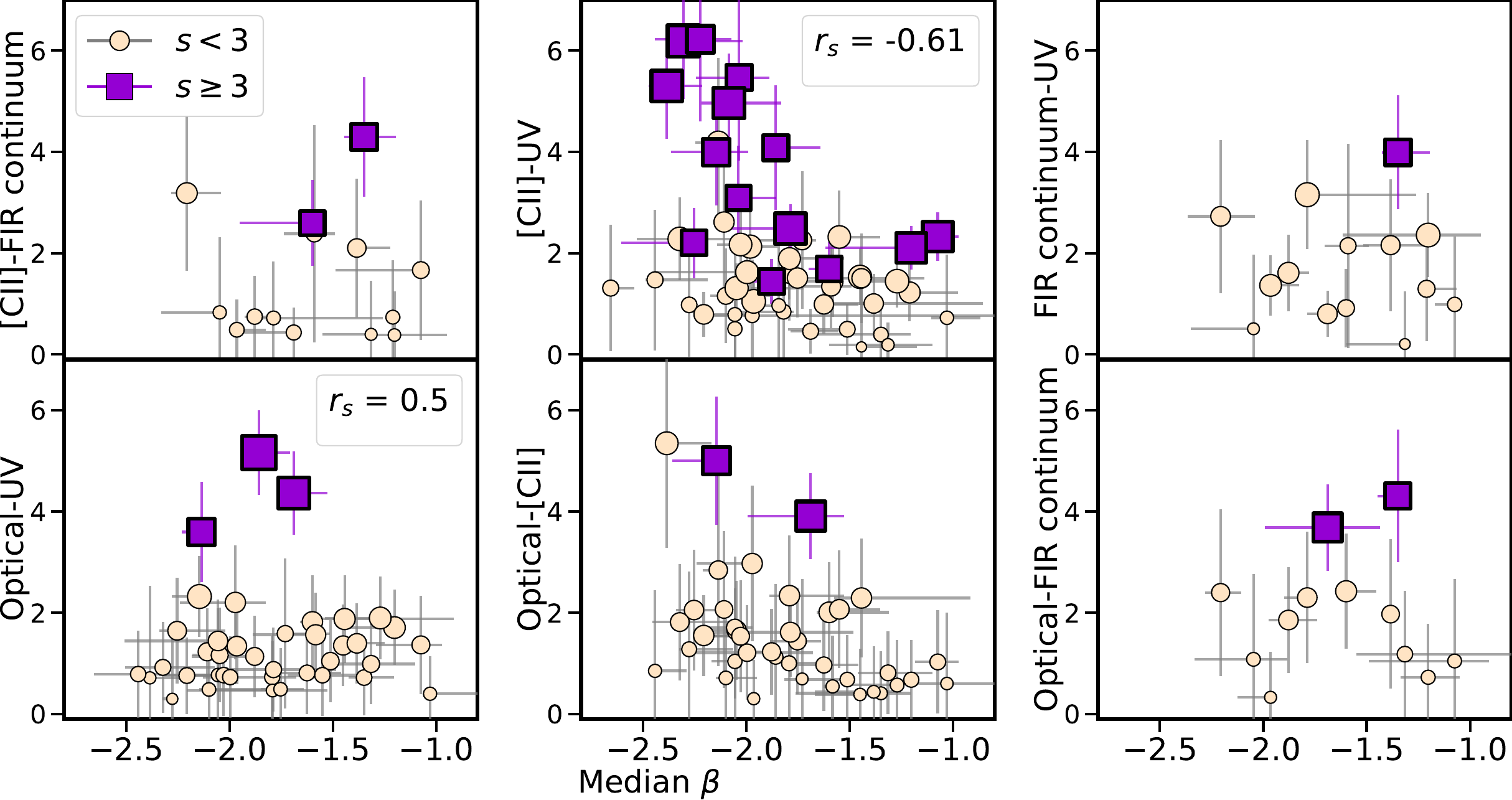}
         \label{fig:multi_beta}
     \end{subfigure}
     \medskip
     \begin{subfigure}[b]{0.95\textwidth}
         \centering
         \includegraphics[width=\textwidth]{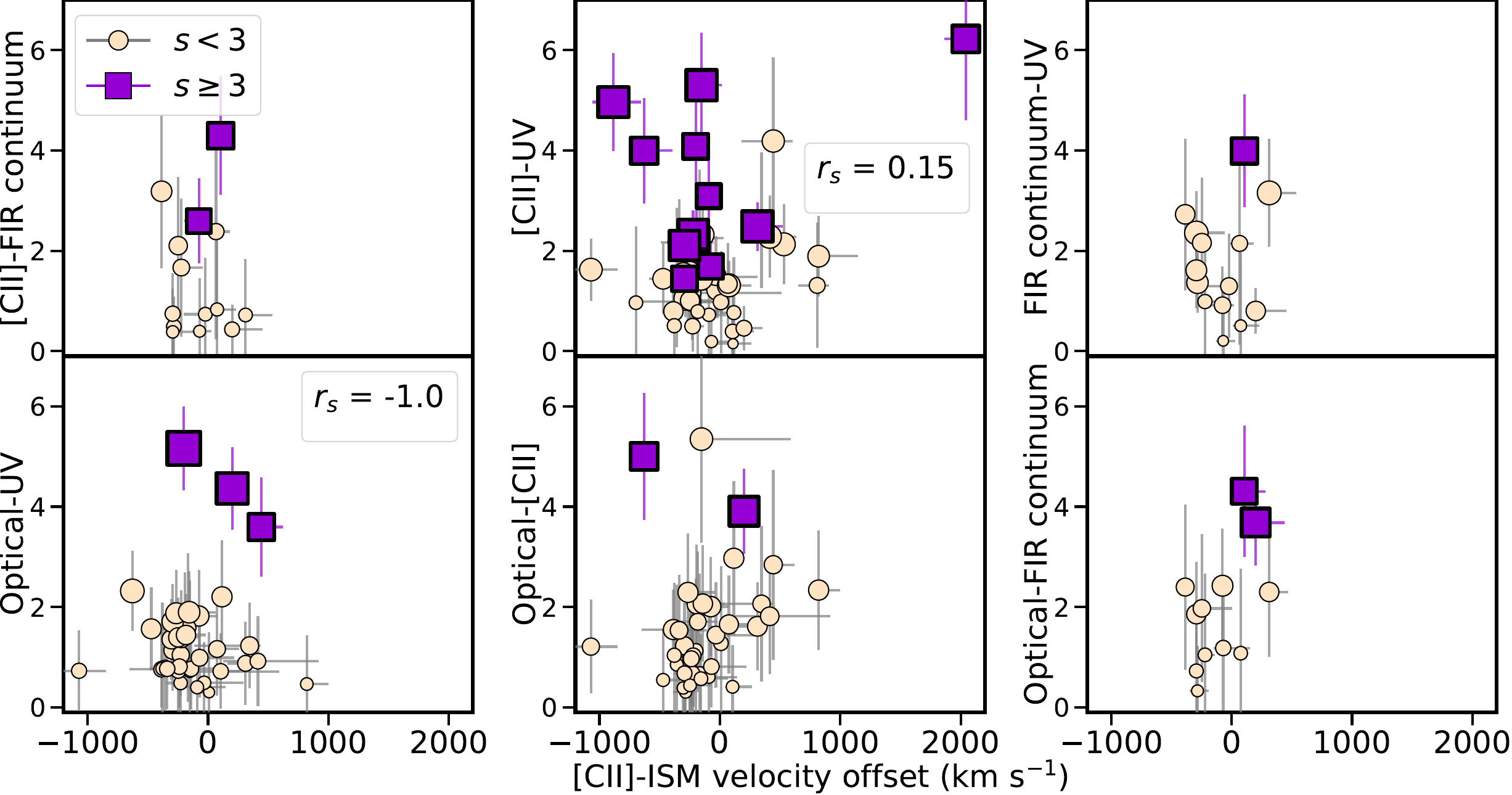}
         \label{fig:multi_vel_off}
     \end{subfigure}
\caption{(contd.) g) median UV continuum slope $\beta$ (as in Fig.~\ref{fig:beta_cont_uv}) h) \lya-\cii velocity offset (as in Fig.~\ref{fig:veloff_cii_uv})}
\label{fig:physical_properties3}
\end{figure*}

% \begin{figure*}
% \ContinuedFloat
%      \begin{subfigure}[b]{0.95\textwidth}
%          \centering
%          \includegraphics[width=\textwidth]{multi_scatter_lya_ew.pdf}
%          \label{fig:multi_lya_ew}
%      \end{subfigure}
%      \medskip
%      \begin{subfigure}[b]{0.95\textwidth}
%          \centering
%          \includegraphics[width=\textwidth]{multi_scatter_f_lya.pdf}
%          \label{fig:multi_flya}
%      \end{subfigure}
% \caption{(contd.) i) (Observer-frame) \lya equivalent width, j) \lya flux}
% \label{fig:physical_properties4}
% \end{figure*}

\begin{figure*}
\ContinuedFloat
     \begin{subfigure}[b]{0.95\textwidth}
         \centering
         \includegraphics[width=\textwidth]{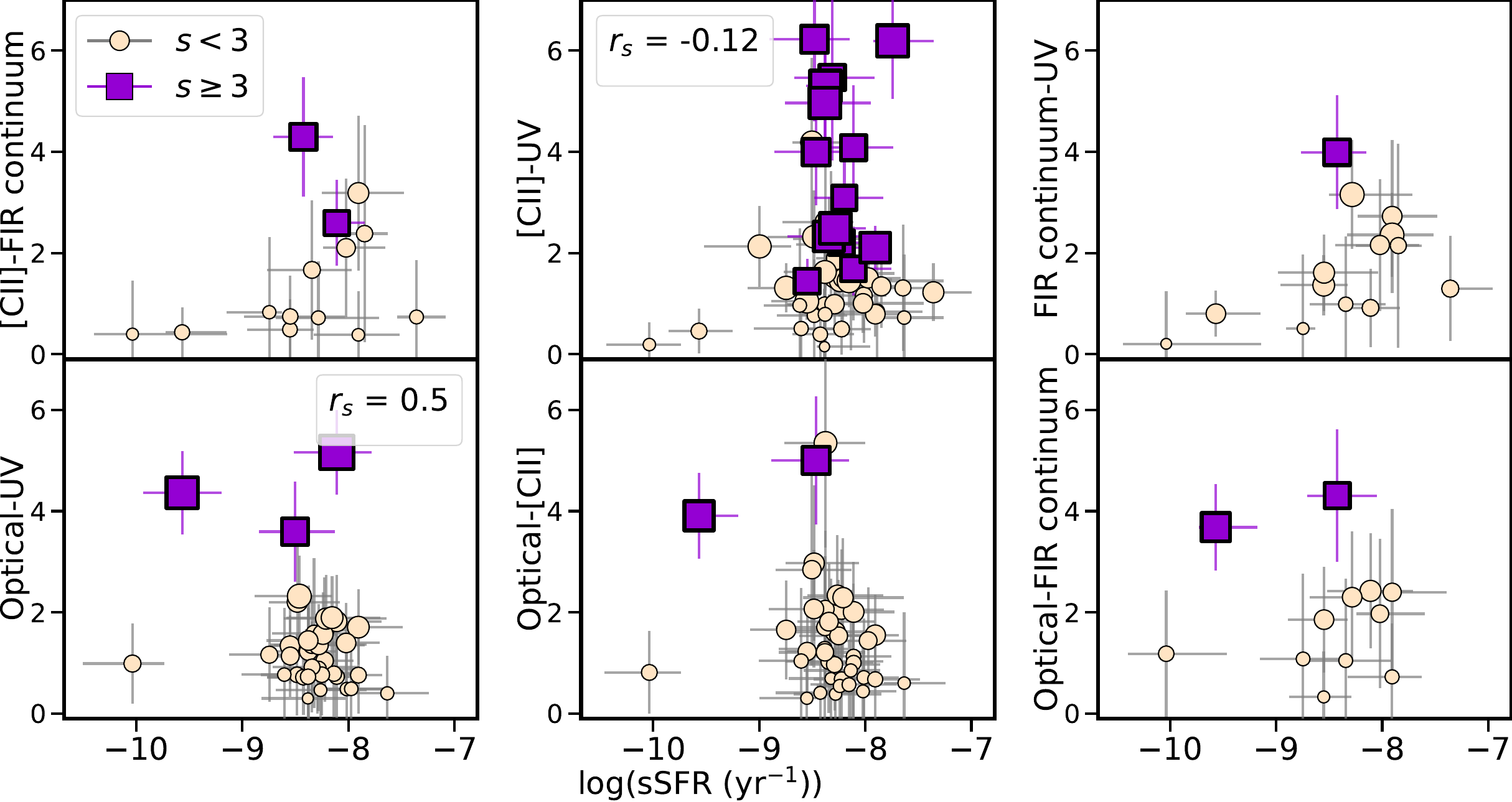}
         \label{fig:multi_ssfr}
     \end{subfigure}
     \medskip
     \begin{subfigure}[b]{0.95\textwidth}
         \centering
         \includegraphics[width=\textwidth]{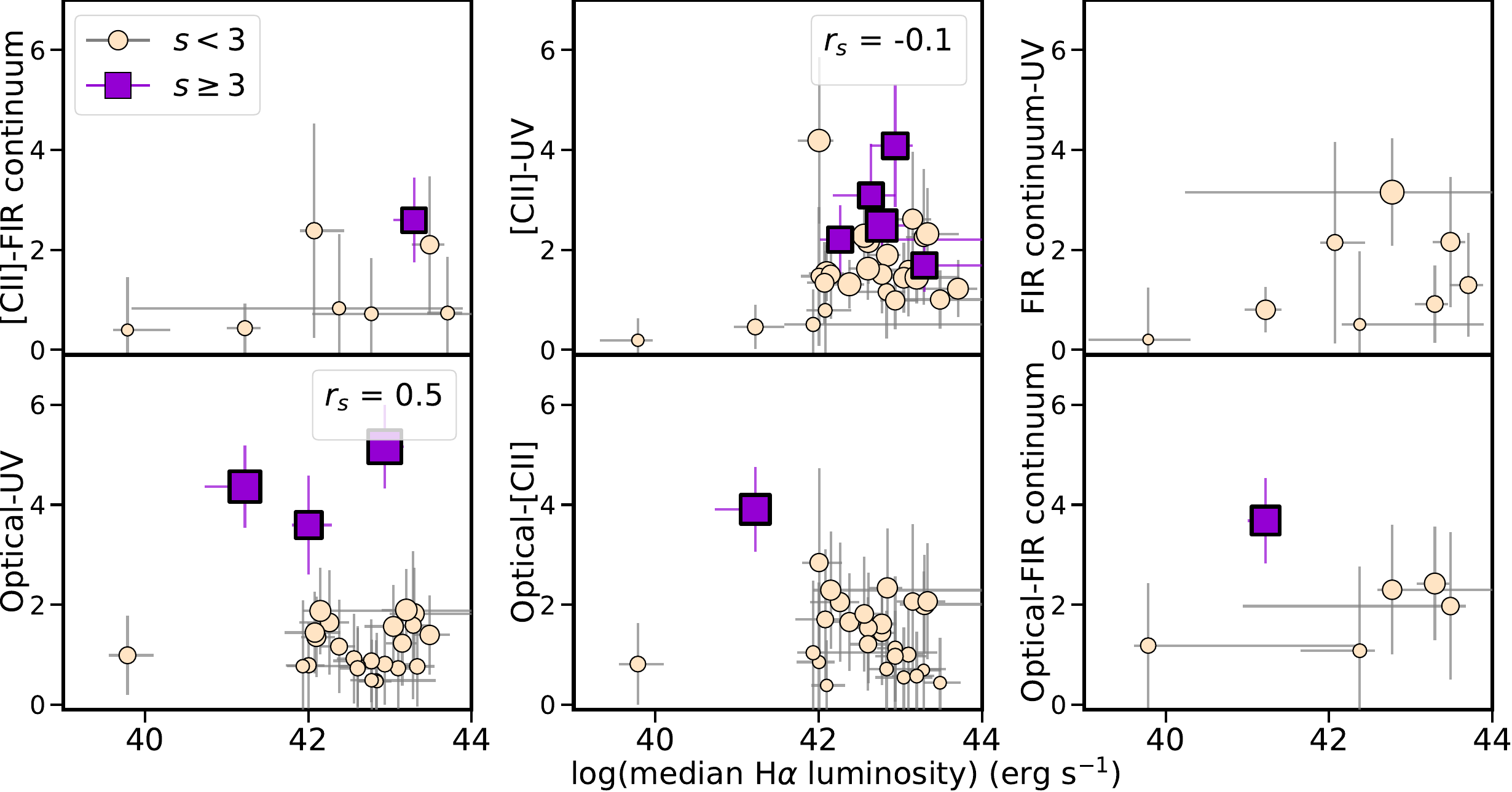}
         \label{fig:multi_halum}
     \end{subfigure}
\caption{(contd.) i) Total specific SFR, j) median \ha luminosity}
\label{fig:physical_properties5}
\end{figure*}

\begin{figure*}
\ContinuedFloat
     \begin{subfigure}[b]{0.95\textwidth}
         \centering
         \includegraphics[width=\textwidth]{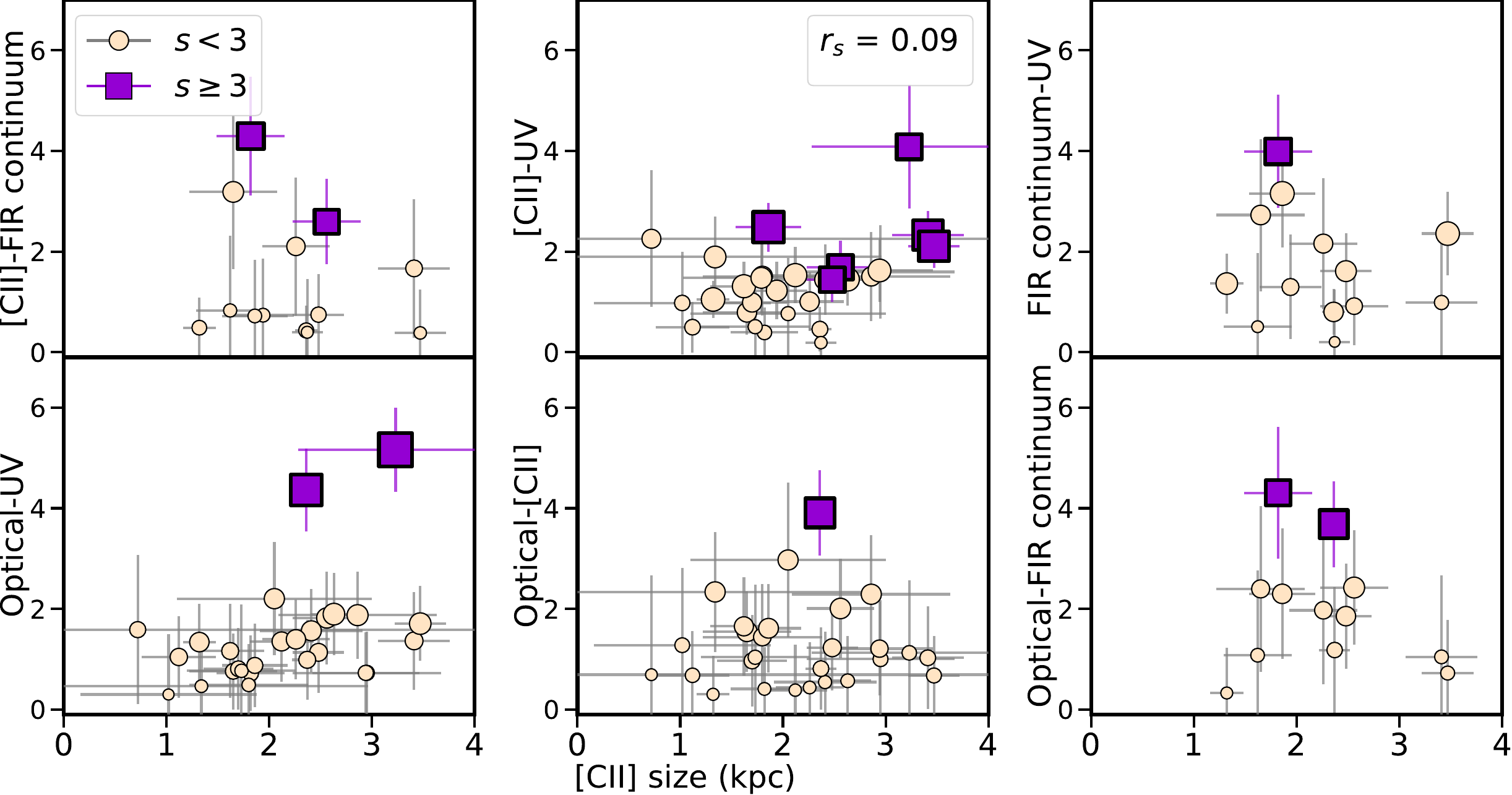}
         \label{fig:multi_ciisize}
     \end{subfigure}
     \medskip
     \begin{subfigure}[b]{0.95\textwidth}
         \centering
         \includegraphics[width=\textwidth]{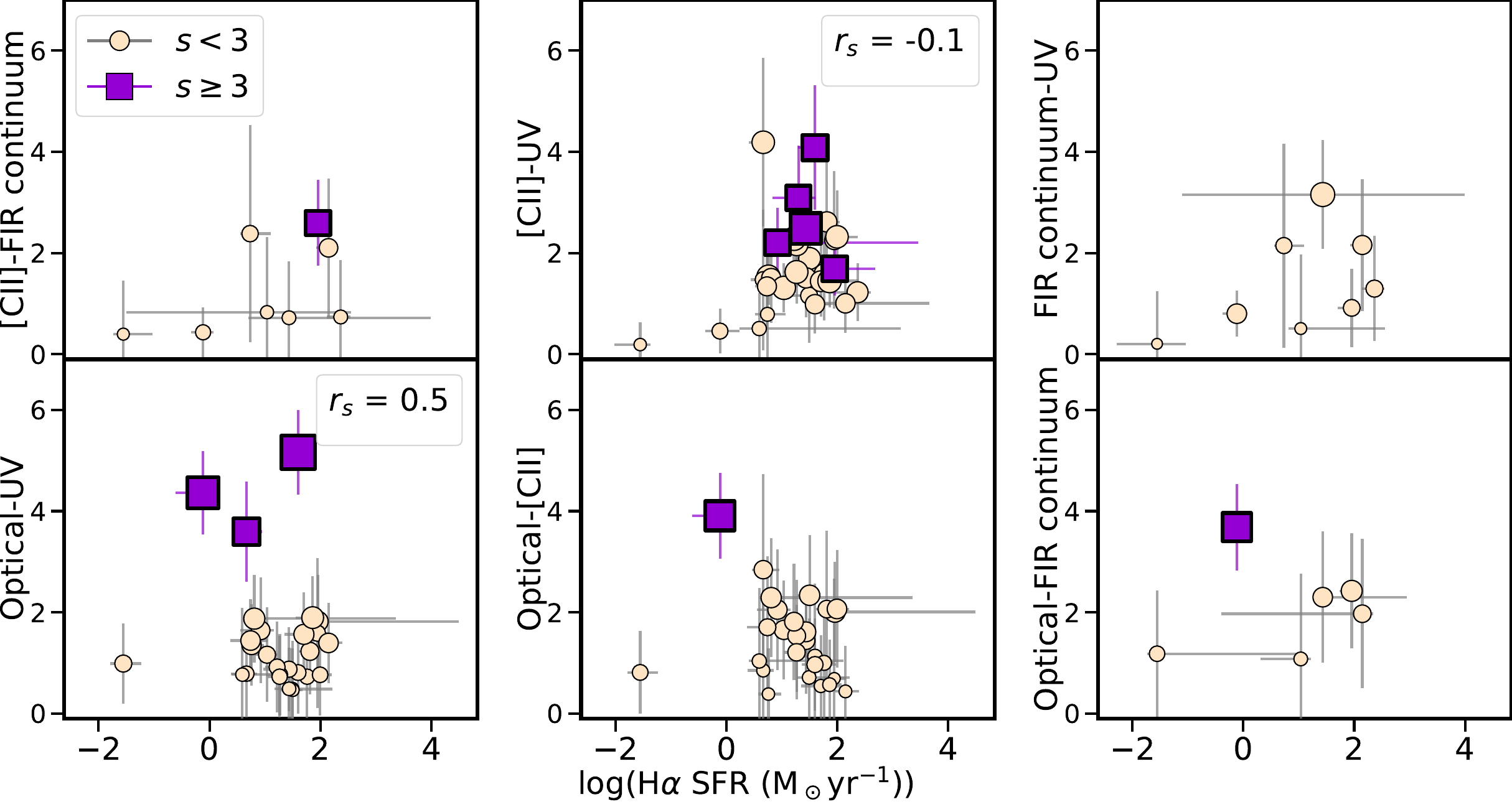}
         \label{fig:multi_sfrha}
     \end{subfigure}
\caption{(contd.) k) \cii size in kpc l) \ha SFR in M$_\odot$ yr$^{-1}$}
\label{fig:physical_properties6}
\end{figure*}

%%%%%%%%%%%%%%%%%%%%%%%%%%%%%%%%%%%%%%%%%%%%%%%%%%

% Don't change these lines
\bsp	% typesetting comment
\label{lastpage}
\end{document}